\documentclass[pdftex]{elsart}
\usepackage{graphicx, amssymb, amsfonts,amsmath}
\usepackage{cite}
\usepackage{subfigure}
\usepackage{hyperref}
\usepackage[bf,footnotesize]{caption}
\usepackage{hyperref}
\usepackage{appendix}
\usepackage{cite}

\graphicspath{{plots/}}

\def \3{\ss }

\newcommand{\Oa}{\mathcal{O}(a)}
\newcommand{\Oasq}{\mathcal{O}(a^2)}

\newcommand{\tr}{\operatorname{Tr}}
\newcommand{\re}{\operatorname{Re}}

\newcommand{\mps}{m_\mathrm{PS}}
\newcommand{\fps}{f_\mathrm{PS}}
\newcommand{\mpcac}{m_\mathrm{PCAC}}

\newcommand{\tauint}{\tau_\mathrm{int}}

\begin{document}

\begin{frontmatter}
  
  \vspace*{-1.0truecm} \title{Light hadrons from lattice QCD with\\
    light ($u,d$), strange and charm\\
    dynamical quarks}
\vspace*{-0.5truecm}
  \begin{center}
\includegraphics[draft=false,width=0.13\linewidth]{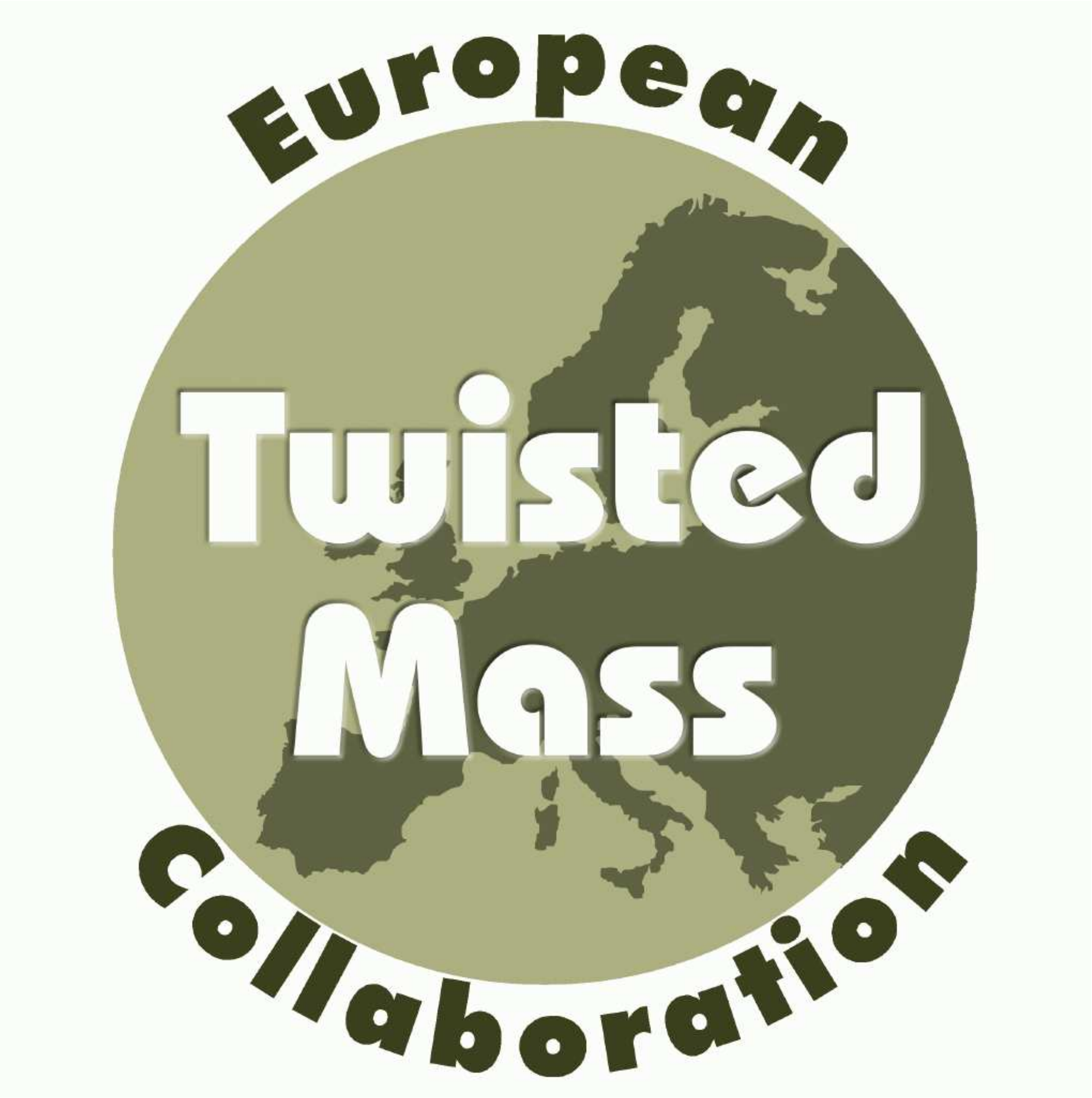}
  \end{center}

  \author{ETM Collaboration},
  \author[a]{R. Baron},
  \author[b]{Ph. Boucaud},
  \author[c]{J. Carbonell},
  \author[d]{A. Deuzeman},
  \author[c]{V. Drach},
  \author[e]{F. Farchioni},
  \author[f]{V. Gimenez},
  \author[g]{G. Herdoiza},
  \author[g]{K. Jansen},
  \author[h]{C. McNeile},
  \author[i]{C. Michael},
  \author[j]{I. Montvay},
  \author[k]{D. Palao},
  \author[d]{E. Pallante},
  \author[b]{O. P\`ene},
  \author[d]{S. Reker},
  \author[l]{C. Urbach},
  \author[m]{M. Wagner},
  \author[n]{U. Wenger}

  \address[a]{ CEA, Centre de Saclay, IRFU/Service de Physique
    Nucl\'eaire, F-91191 Gif-sur-Yvette, France}
  \address[b]{ Laboratoire de Physique Th\'eorique (B\^at. 210),
    CNRS et Universit\'e Paris-Sud 11, Centre d'Orsay, 91405 Orsay-Cedex,
    France}
 \address[c]{ Laboratoire de Physique Subatomique et Cosmologie,
    53 avenue des Martyrs, 38026 Grenoble, France}
  \address[d]{ Centre for Theoretical Physics, University of
    Groningen, Nijenborgh 4, 9747 AG Groningen, the Netherlands}
  \address[e]{Institut f\"ur
    Theoretische Physik, Universit\"at M\"unster, Wilhelm-Klemm-Stra\ss e 9, D-48149 M\"unster,
    Germany}
  \address[f]{ Dep. de F\'isica Te\`orica and IFIC, Universitat
    de Val\`encia-CSIC, Dr.Moliner 50, E-46100 Burjassot, Spain}
  \address[g]{ NIC, DESY, Platanenallee 6, D-15738 Zeuthen,
    Germany}
 \address[h]{ Department of Physics and Astronomy, The Kelvin Building, University of
    Glasgow, G12 8QQ Glasgow, United Kingdom}
  \address[i]{ Division of Theoretical Physics, University of
    Liverpool, L69 3BX Liverpool, United Kingdom}
  \address[j]{ DESY, Notkestr.
    85, D-22603 Hamburg, Germany}
  \address[k]{ INFN, Sez. di Roma "Tor Vergata", Via della Ricerca Scientifica 1, I-00133
Rome, Italy}
  \address[l]{ Helmholtz-Institut f{\"u}r Strahlen- und
    Kernphysik (Theorie) and Bethe Center for Theoretical Physics,
    Universit{\"a}t Bonn, 53115 Bonn, Germany}
  \address[m]{Institut f\"ur Physik, Humboldt-Universit\"at zu Berlin, 
Newtonstra\ss e 15, D-12489 Berlin, Germany}
  \address[n]{ Albert Einstein Center for Fundamental Physics,
    Institute for Theoretical Physics, University of Bern, Sidlerstr.
    5, CH-3012 Bern, Switzerland}
  \clearpage

  \begin{abstract}

    \noindent We present results of lattice QCD simulations with
    mass-degenerate up and down and mass-split strange and charm
    ($N_{\rm f} = 2+1+1$) dynamical quarks using Wilson twisted mass
    fermions at maximal twist. The tuning of the strange and charm
    quark masses is performed at two values of the lattice
    spacing $a\approx 0.078$\,fm and $a\approx 0.086$\,fm with lattice
    sizes ranging from { $L\approx 1.9$\,fm} to { $L\approx 2.8$\,fm}.
    We measure with high statistical precision the light pseudoscalar
    mass $m_\mathrm{PS}$ and decay constant $f_\mathrm{PS}$ in a range
    $270\lesssim m_\mathrm{PS}\lesssim 510$\,MeV and determine the low
    energy parameters $f_0$ and $\bar{l}_{3,4}$ of SU(2) chiral
    perturbation theory.  We use the two values of the lattice
    spacing, several lattice sizes as well as different values of the
    light, strange and charm quark masses to explore the systematic
    effects. A first study of discretisation effects in light-quark
    observables and a comparison to $N_{\rm f}=2$ results are
    performed.

  \end{abstract}

  \begin{keyword}
    Lattice gauge theory, lattice QCD, light hadrons, charm quark,
    chiral perturbation theory.
    \PACS 12.38.Gc \sep 12.39Fe\\
    Preprint-No: DESY 10-054, HU-EP-10/18, IFIC/10-11, SFB/CPP-10-29, LPT-Orsay 10-28, LTH873, LPSC1042, MS-TP-10-09, ROM2F/2010/08 \\
  \end{keyword}

\end{frontmatter}


\section{Introduction and Main Results} 

The beginning of this century has assisted to radical improvements in
theory, algorithms and supercomputer technology, leading to a far
increased ability to solve non-perturbative aspects of gauge field
theories in a lattice regularised framework. Following this path of
improving the lattice setup, in this paper, we are reporting about our
experiences and results when considering in addition to the $u,d$
light dynamical flavours also the effects of the strange and charm sea
quarks. By including a dynamical charm, we are now able to directly
study its contribution to physical observables and to quantify the so
far uncontrolled systematic effect present in lattice QCD simulations
where the charm flavour in the sea is absent.

A number of different lattice fermion formulations are being used by
several lattice groups, see refs.~\cite{Jansen:2008vs,Scholz:2009yz}
for recent reviews.  Here, we adopt a particular type of Wilson
fermions, known as the Wilson twisted mass formulation of lattice QCD
(tmLQCD), introduced in~\cite{Frezzotti:2000nk,Frezzotti:2003ni}.
This approach is by now well established, with many physical results
obtained with two light degenerate twisted mass flavours ($N_{\rm
  f}=2$) by our European Twisted Mass (ETM) Collaboration, see
refs.~\cite{Boucaud:2007uk,Blossier:2007vv,Urbach:2007rt,
  Boucaud:2008xu,Cichy:2008gk,Alexandrou:2008tn,Jansen:2008si,Jansen:2008wv,Frezzotti:2008dr,Blossier:2009bx,Jansen:2009hr,McNeile:2009mx,
  Baron:2009wt,Alexandrou:2009qu,Blossier:2009hg,Blossier:2009vy,Constantinou:2010gr,Michael:2010iv}.
For a review see ref.~\cite{Shindler:2007vp}. In the tmLQCD
formulation a twisted mass term is added to the standard, unimproved
Wilson-Dirac operator, and the formulation becomes especially
interesting when the theory is tuned to maximal
twist~\cite{Frezzotti:2003ni}.  The major advantage of the lattice
theory tuned to maximal twist is the automatic $\Oa$ improvement of
physical observables, independently of the specific type of operator
considered, implying that no additional, operator specific improvement
coefficients need to be computed. Other advantages worth to mention
are that the twisted mass term acts as an infrared regulator of the
theory and that mixing patterns in the renormalisation procedure are
expected to be simplified.

Detailed studies of the continuum-limit scaling in the quenched
approximation~\cite{Jansen:2003ir,Jansen:2005gf,Jansen:2005kk,Abdel-Rehim:2005gz}
and with two dynamical
quarks~\cite{Baron:2009wt,Urbach:2007rt,Dimopoulos:2007qy,Alexandrou:2008tn}
have demonstrated that, after an appropriate tuning procedure
to maximal twist, lattice artefacts not only follow the expected $\Oasq$ scaling
behaviour~\cite{Frezzotti:2003ni}, but also that the remaining $\Oasq$
effects are small, in agreement with the conclusions drawn in
ref.~\cite{Frezzotti:2005gi}.

The only exception seen so far is the neutral pseudoscalar mass, which
shows significant $\Oasq$ effects. This arises from the explicit
breaking of both parity and isospin symmetry, which are however
restored in the continuum limit with a rate of $\Oasq$ as shown
in~\cite{Frezzotti:2003ni} and numerically confirmed in
refs.~\cite{Jansen:2005cg,Baron:2009wt}. Moreover, a recent analysis
suggests that isospin breaking effects strongly affect only a limited
set of observables, namely the neutral pion mass and kinematically
related quantities~\cite{Frezzotti:2007qv,Dimopoulos:2009qv}.

In this paper we report on simulations with twisted mass dynamical up,
down, strange and charm quarks. We realise this by adding a heavy
mass-split doublet $(c,s)$ to the light degenerate mass doublet
$(u,d)$, referring to this setup as $N_{\rm f}=2+1+1$ simulations.
This formulation was introduced
in~\cite{Frezzotti:2004wz,Frezzotti:2003xj} and first explored
in~\cite{Chiarappa:2006ae}. As for the mass-degenerate case, the use
of lattice action symmetries allows to prove the automatic $\Oa$
improvement of physical observables in the non-degenerate
case~\cite{Frezzotti:2004wz,Frezzotti:2003xj}. First accounts of our
work were presented at recent
conferences~\cite{Baron:2008xa,Baron:2009zq}. Recently, results with
$N_{\rm f}=2+1+1$ staggered fermions have been reported
in~\cite{Bazavov:2009jc,Bazavov:2009wm,Bazavov:2010ru}.  The inclusion
of the strange and charm degrees of freedom allows for a most complete
description of light hadron physics and eventually opens the way to
explore effects of a dynamical charm in genuinely strong interaction
processes and in weak matrix elements.

\begin{figure}[t]
  \centering
    \subfigure[\label{fig:amps2_vs_amu}]%
  {\includegraphics[width=0.47\linewidth]{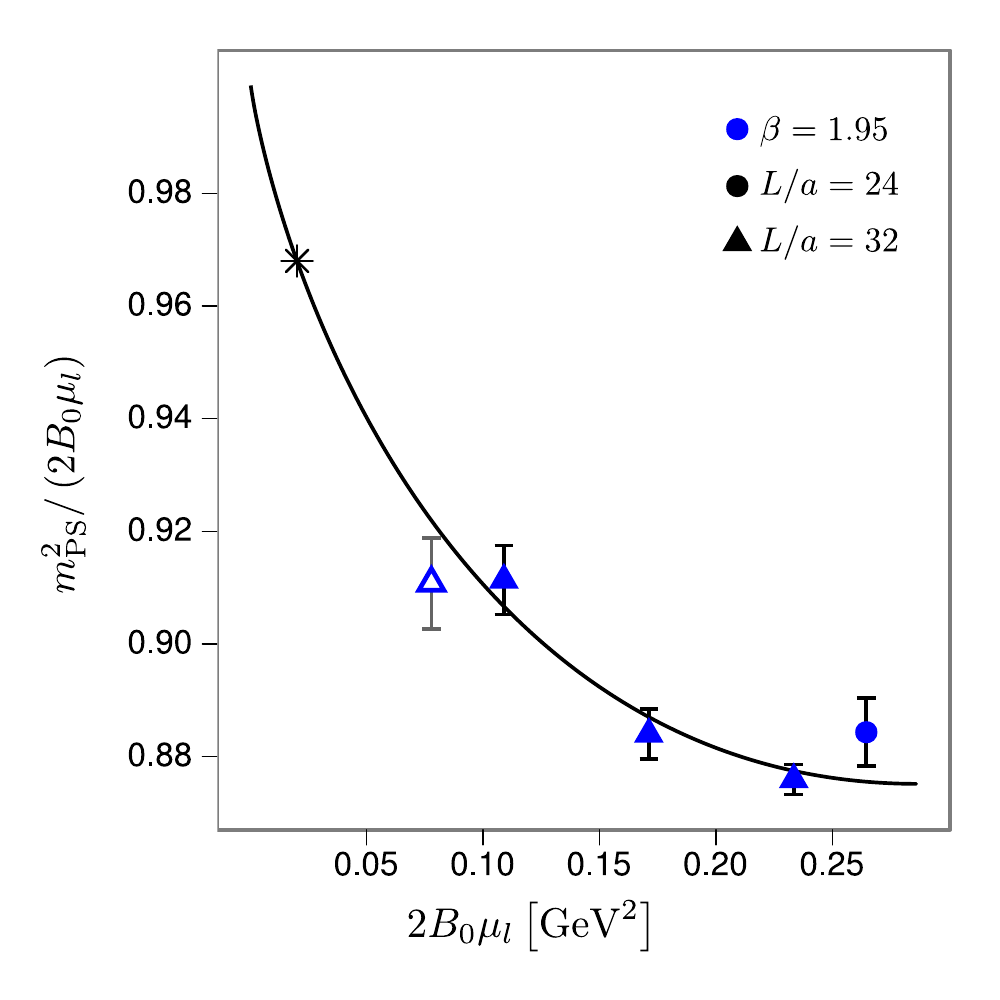}}
  \subfigure[\label{fig:afps_vs_amu}]%
  {\includegraphics[width=0.47\linewidth]{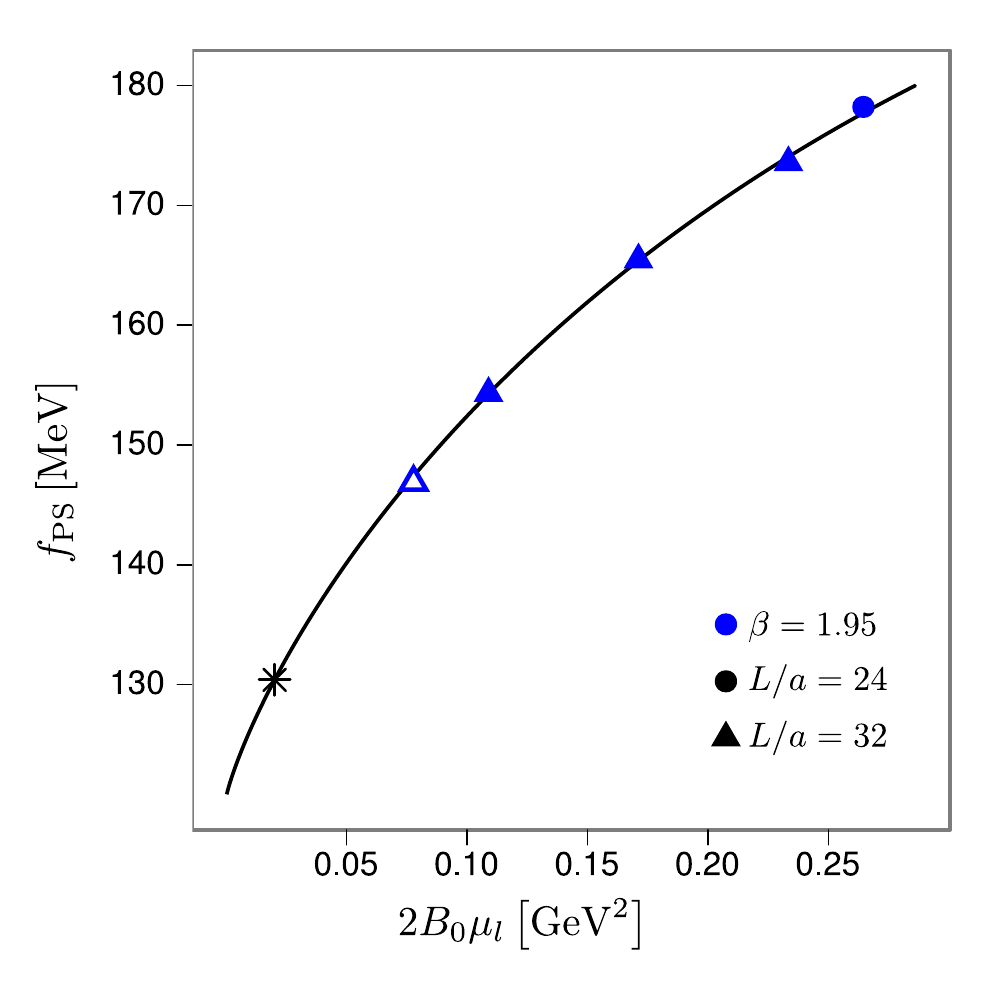}}
  \caption{(a) The charged pseudoscalar mass ratio
    $\mps^2/(2B_0\mu_l)$ and (b) the pseudoscalar decay constant
    $\fps$ as a function of $2B_0\mu_l$ fitted to
    SU(2) chiral perturbation theory, see
    table~\ref{table:fitresults}.  
The scale is set by the value of
$2B_0\mu_{l}$ at which the ratio
$\fps^{[L=\infty]}/\mps^{[L=\infty]}$ assumes its physical
value~\cite{Amsler:2008zzb} $f_\pi/m_\pi = 130.4(2)/135.0$ (black star).
The lattice gauge coupling is $\beta=1.95$
    and the twisted light quark mass ranges from $a\mu_l = 0.0025$ to
    $0.0085$, see eq. (\ref{eq:sl}) for its definition, corresponding to a range of 
the pseudoscalar mass $270\lesssim \mps \lesssim 490$\,MeV. 
The kaon and $D$ meson masses are
      tuned to their physical value, see
    table~\ref{tab:kdmasses}. The lightest point (open symbol) has
    not been included in the chiral fit, see the discussion in
    section~\ref{sec:tuning}.}
  \label{fig:chiralfit}
\end{figure}

Here, we concentrate on results in the light-quark sector using the
charged pseudoscalar mass $\mps$ and decay constant $\fps$ as basic
observables involving up and down valence quarks only. In
fig.~\ref{fig:chiralfit} we show the dependence of (a)
$\mps^2/2B_0\mu_l$ and (b) $\fps$ as a function of the mass parameter
$2B_0\mu_l$, together with a fit to SU(2)
chiral perturbation theory ($\chi$PT) at the smallest value of the lattice
spacing of $a\approx 0.078\,{\mbox{fm}}$ and lattice gauge coupling $\beta =1.95$.  
We summarise the fit results for
the low energy constants in table~\ref{table:fitresults}. These are the main results 
of this paper. 

 A comparison between data obtained with $N_{\rm f}=2+1+1$ and
  $N_{\rm f}=2$ flavours of quarks - see sections~\ref{sec:discr} and 
\ref{sec:observables}, and ref.~\cite{Baron:2009wt} - reveals a
  remarkable agreement for the results involving light-quark
  observables such as the pseudoscalar mass and decay constant or the
  nucleon mass. This provides a strong indication in favour of the
good quality of our data in this new setup. In particular, barring cancellations 
due to lattice discretisation errors, these results
would suggest that the dynamical strange and charm degrees of freedom do
not induce large effects in these light-quark observables.  In the
$N_{\rm f}=2$ case, data collected at four values of the lattice
spacing have allowed us to properly quantify all systematic errors
present in the determination of light-quark
observables~\cite{Baron:2009wt}.  In this first work with $N_{\rm
  f}=2+1+1$ flavours, we consider data at two close values of the
lattice spacing, while we defer to a forthcoming publication the
inclusion of additional ensembles at a significantly lower lattice
spacing and a more complete analysis of the systematic effects.

\begin{table}[t!]
  \centering
  \begin{tabular*}{.60\linewidth}{@{\extracolsep{\fill}}lr}
    \hline\hline
    $\Bigl.\Bigr.$ & $\beta=1.95$ \\
    \hline\hline
    $\bar{l}_3$             &  3.70(7)(26)\\
    $\bar{l}_4$             &  4.67(3)(10)\\
    $f_0\ [\mathrm{MeV}]$   &  121.14(8)(19)\\
    $f_\pi/f_0$             &  1.076(2)(2)\\
    $2B_0\mu_{u,d}/m_\pi^2$ &  1.032(21)(3)\\
    $\langle r^2\rangle_s^\mathrm{NLO}\ [\mathrm{fm}^2]$
                            &  0.724(5)(23)\\
    \hline
    $r^\chi_0/a(\beta=1.95)$& 5.71(4)  \\
    $r^\chi_0(\beta=1.95)\ [\mathrm{fm}]$ & 0.447(5)\\
    $a(\beta=1.95)\ [\mathrm{fm}]$        & 0.0782(6)\\
    \hline
  \end{tabular*}
  \caption{Results of the fits to SU(2) $\chi$PT for the ensemble at
    $\beta=1.95$. Predicted quantities are: the low energy constants
    $\bar{l}_{3,4}$, the charged pseudoscalar decay constant in the
    chiral limit $f_0$, the mass ratio $2B_0\mu_l/\mps^2$ at the
    physical point and the pion scalar radius $\langle
    r^2\rangle_s^\mathrm{NLO}$. The first quoted error is from the
    chiral fit at $\beta =1.95$, the second error is the systematic
    uncertainty that conservatively accommodates the best fitted
    central values of the three fits reported in table
    \ref{table:fitresults_c}, section \ref{sec:observables}. The small
    error on the quoted lattice spacing comes exclusively from the fit
    at $\beta =1.95$.  The scale is set by fixing the ratio
    $\fps^{[L=\infty]}/\mps^{[L=\infty]} = f_\pi/m_\pi = 130.4(2)/135.0$ to its physical
    value~\cite{Amsler:2008zzb}. The chirally extrapolated Sommer
    scale $r_0^\chi$ is determined separately and not included in
    the $\chi$PT fits.  For a comparison with the $N_{\rm f}=2$ ETMC
    results, see~\cite{Baron:2009wt}. }
  \label{table:fitresults}
\end{table}

The rest of this paper is organised as follows.  In
section~\ref{sec:setup} we describe the gauge action and the twisted
mass fermionic action for the light and heavy sectors of the
theory. The realisation of $\Oa$ improvement at maximal twist is also
presented. In section~\ref{sec:simdet} we define the simulation
parameters, describe the tuning to maximal twist as well as the tuning
of the strange and charm quark masses and the relevance of
discretisation effects. Section~\ref{sec:observables} includes a
discussion of the fits to SU(2) $\chi$PT also for data on a slightly
coarser lattice, $a\approx 0.086\,{\mbox{fm}}$, and provides a first
account of systematic uncertainties. Our conclusions and future
prospects are summarised in section~\ref{sec:concl}.

\section{Lattice Action}
\label{sec:setup}

The complete lattice action can be written as
\begin{equation}
  S= S_g + S_l + S_h\, ,
\end{equation}
where $S_g$ is the pure gauge action, in our case the so-called
Iwasaki action~\cite{Iwasaki:1985we,Iwasaki:1996sn}, $S_l$ is the twisted mass
Wilson action for the light
doublet~\cite{Frezzotti:2000nk,Frezzotti:2003ni} and $S_h$ the one for
the heavy doublet~\cite{Frezzotti:2004wz,Frezzotti:2003xj}.

\subsection{Gauge action}
\label{sec:gauge}

The Iwasaki gauge action~\cite{Iwasaki:1985we,Iwasaki:1996sn} includes besides the
plaquette term $U^{1\times1}_{x,\mu,\nu}$ also rectangular
$(1\times2)$ Wilson loops $U^{1\times2}_{x,\mu,\nu}$
\begin{equation}
  \label{eq:Sg}
    S_g =  \frac{\beta}{3}\sum_x\Biggl(  b_0\sum_{\substack{
      \mu,\nu=1\\1\leq\mu<\nu}}^4\{1-\re\tr(U^{1\times1}_{x,\mu,\nu})\}\Bigr. 
     \Bigl.+
    b_1\sum_{\substack{\mu,\nu=1\\\mu\neq\nu}}^4\{1
    -\re\tr(U^{1\times2}_{x,\mu,\nu})\}\Biggr)\, ,
\end{equation}
with $\beta=6/g_0^2$ the bare inverse coupling, $b_1=-0.331$ and the
 normalisation condition $b_0=1-8b_1$.

The choice of the gauge action is motivated by the non trivial
phase structure of Wilson-type fermions at finite values of the
lattice spacing. The phase structure of the theory has been
extensively studied analytically, by means of chiral perturbation
theory~\cite{Munster:2003ba,Sharpe:2004ny,Aoki:2004ta,Sharpe:2004ps,Munster:2004am,Munster:2004wt,Scorzato:2004da},
and
numerically~\cite{Farchioni:2004us,Farchioni:2004ma,Farchioni:2004fs,Farchioni:2005ec,Farchioni:2005bh,Farchioni:2005tu}.
These studies provided evidence for a first order phase transition
close to the chiral point for coarse lattices. 
This implies that simulations at non-vanishing lattice spacing 
cannot be performed with pseudoscalar masses below a
minimal critical value.

The strength of the phase transition has been found
\cite{Farchioni:2004fs,Farchioni:2005tu} to be highly sensitive to the
value of the parameter $b_1$ in the gauge
action in eq.~(\ref{eq:Sg}). Moreover, in~\cite{Chiarappa:2006ae} it was
observed that its strength grows when increasing the number of
flavours in the sea from $N_{\rm f}=2$ to $N_{\rm f}=2+1+1$, at
otherwise fixed physical situation.  Numerical studies with our
$N_{\rm f}=2+1+1$ setup have shown that the Iwasaki gauge action, with
$b_1=-0.331$, provides a smoother dependence of phase transition
sensitive quantities on the bare quark mass than the
tree-level-improved Symanzik~\cite{Weisz:1982zw,Weisz:1983bn} gauge
action, with $b_1=-1/12$, chosen for our $N_{\rm f}=2$ simulations.

Another way to weaken the strength of the phase transition is to
modify the covariant derivative in the fermion action by smearing the
gauge fields. While the main results of this work do not use smearing of 
the gauge fields, we report in
section~\ref{sec:stout} on our experience 
when applying a stout smearing~\cite{Morningstar:2003gk} procedure,
see also~\cite{Jansen:2007sr}.

\subsection{Action for the Light Doublet}

The lattice action for the mass degenerate light doublet $(u,d)$ in
the so called twisted basis
reads~\cite{Frezzotti:2000nk,Frezzotti:2003ni}
\begin{equation}
  \label{eq:sl}
  S_l\ =\ a^4\sum_x\left\{ \bar\chi_l(x)\left[ D[U] + m_{0,l} +
    i\mu_l\gamma_5\tau_3\right]\chi_l(x)\right\}\, ,
\end{equation}
where $m_{0,l}$ is the untwisted bare quark mass, $\mu_l$ is the bare
twisted light quark mass, $\tau_3$ is the third Pauli matrix acting in
flavour space and
\[
D[U] = \frac{1}{2}\left[\gamma_\mu\left(\nabla_\mu +
  \nabla^*_\mu\right) -a\nabla^*_\mu\nabla_\mu \right]
\]
is the massless Wilson-Dirac operator. $\nabla_\mu$ and
$\nabla^*_\mu$ are the forward and backward gauge covariant difference
operators, respectively.  Twisted mass light fermions are said to be
at maximal twist if the bare untwisted mass $m_{0,l}$ is tuned to its
critical value, $m_{\rm crit}$, the situation we shall reproduce in
our simulations. The quark doublet $\chi_l=(\chi_u, \chi_d)$ in the
twisted basis is related by a chiral rotation to the quark doublet in
the physical basis
\begin{equation}
  \psi_{l}^{phys} = e^{\frac{i}{2}\omega_{l}\gamma_{5}\tau_{3}}\chi_{l}, 
  \qquad \bar{\psi}_{l}^{phys}=\bar{\chi}_{l}e^{\frac{i}{2}\omega_{l}\gamma_{5}\tau_{3}}\, ,
\end{equation}
where the twisting angle $\omega_{l}$ takes the value $|\omega_{l}| \to
\frac{\pi}{2}$ as $|m_{0,l} - m_{\rm crit}| \to 0$. We shall use the twisted basis 
throughout this paper.

\subsection{Action for the Heavy Doublet}

We introduce a dynamical strange quark by adding a twisted heavy
mass-split doublet $\chi_h = (\chi_c, \chi_s)$, thus also introducing
a dynamical charm in our framework.  As shown in
\cite{Frezzotti:2003xj}, a real quark determinant can in this case be
obtained if the mass splitting is taken to be orthogonal in isospin
space to the twist direction.  We thus choose the
construction~\cite{Frezzotti:2004wz,Frezzotti:2003xj}
\begin{equation}
  \label{eq:sf}
  S_h\ =\ a^4\sum_x\left\{ \bar\chi_h(x)\left[ D[U] + m_{0,h} +
    i\mu_\sigma\gamma_5\tau_1 + \mu_\delta \tau_3 \right]\chi_h(x)\right\}\, ,
\end{equation}
where $m_{0,h}$ is the untwisted bare quark mass for the heavy
doublet, $\mu_\sigma$ the bare twisted mass --
the twist is this time along the $\tau_{1}$ direction -- and
$\mu_\delta$ the mass splitting along the $\tau_{3}$ direction.

The bare mass parameters $\mu_\sigma$ and $\mu_\delta$ of the
non-degenerate heavy doublet are  related to the physical renormalised strange and charm quark masses
via~\cite{Frezzotti:2004wz}
\begin{equation}
  \begin{split}
    \label{eq:msmc}
    (m_{s})_{\rm R} &= Z_{\rm P}^{-1} \, (\mu_\sigma - Z_{\rm P}/Z_{\rm S} \, \mu_\delta)\,, \\
    (m_{c})_{\rm R} &= Z_{\rm P}^{-1} \, (\mu_\sigma + Z_{\rm P}/Z_{\rm S} \, \mu_\delta)\, ,
  \end{split}
\end{equation}
where $Z_{\rm P}$ and $Z_{\rm S}$ are the renormalisation constants of
the pseudoscalar and scalar quark densities, respectively, computed in the massless
standard Wilson theory.

A chiral rotation analogous to the one in the light sector transforms
the heavy quark doublet from the twisted to the physical basis
\begin{equation}
  \psi_{h}^{phys} = e^{\frac{i}{2}\omega_{h}\gamma_{5}\tau_{1}}\chi_{h},
  \qquad \bar{\psi}_{h}^{phys}=\bar{\chi}_{h}e^{\frac{i}{2}\omega_{h}\gamma_{5}\tau_{1}},
\end{equation}
where the twisting angle $\omega_{h}$ takes the value $|\omega_{h}|
\to \frac{\pi}{2}$ as $|m_{0,h} - m_{\rm crit}| \to 0$.
\subsection{$\Oa$ improvement at maximal twist}
\label{sec:maxtwist}

One of the main advantages of Wilson twisted mass fermions is that by
tuning the untwisted bare quark mass to its critical value, automatic
$\Oa$ improvement of physical observables can be achieved.

Tuning the complete $N_{\rm f}=2+1+1$ action to maximal twist can in
principle be performed by independently choosing the bare masses of
the light and heavy sectors $am_{0,l}$ and $am_{0,h}$, resulting,
however, in a quite demanding procedure. On the other
hand, properties of the Wilson twisted mass formulation allow for a
rather economical, while accurate alternative
\cite{Frezzotti:2003ni,Frezzotti:2003xj,Chiarappa:2006ae}, where the
choice $am_{0,l}=am_{0,h}\equiv {1}/{2\kappa} -4$ is made, and the
hopping parameter $\kappa$ has been introduced.

Tuning to maximal twist, i.e. $\kappa =\kappa_{crit}$, is then
achieved by choosing a parity odd operator $O$ and determine
$am_{crit}$ (equivalently $\kappa_{crit}$) such that $O$ has vanishing
expectation value. One appropriate quantity is the PCAC light quark mass
\cite{Farchioni:2004ma,Farchioni:2004fs,Frezzotti:2005gi}
\begin{equation}
  \label{eq:pcacmass}
  \mpcac = \frac{\sum_{{\bf x}}\left< \partial_{0}A^{a}_{0,l}({\bf x},t)P_{l}^{a}(0)\right>}
         {2\sum_{{\bf x}}\left<P^{a}_{l}({\bf x},t)P^{a}_{l}(0)\right>},\qquad a=1,2 \; ,
\end{equation}
where
\begin{equation}
  A_{\mu,l}^a(x) = \bar{\chi_l} (x)\gamma_\mu\gamma_5\frac{\tau_a}{2}\chi_l (x)\, ,~~~~~~~~~~~~~
  P_l^a(x) = \bar{\chi_l} (x)\gamma_5\frac{\tau_a}{2}\chi_l (x)\, , 
\end{equation}
and we demand $\mpcac =0$.
 For the quenched~\cite{Jansen:2005gf} and the $N_{\rm f}=2$ case
~\cite{Baron:2009wt}, this method has been found to be successful in providing the 
expected $\Oa$ improvement and effectively 
 reducing residual $\Oasq$ discretisation effects in the region of small
quark masses~\cite{Frezzotti:2005gi}.

The numerical precision required for the tuning of $\mpcac$ to zero
has been discussed in~\cite{Boucaud:2008xu}. Contrary to the $N_{\rm
  f}=2$ case~\cite{Boucaud:2007uk,Boucaud:2008xu}, where this tuning
was performed once at the minimal value of the twisted light mass
considered in the simulations, we now perform the tuning at each value
of the twisted light quark mass $\mu_l$ and the heavy-doublet quark 
mass parameters $\mu_\sigma$ and $\mu_\delta$. This obviously leaves more
freedom in the choice of light quark masses for future computations.

Although theoretical arguments tell us that $\Oa$ improvement
is at work in our setup, a dedicated continuum scaling study is always 
required to accurately quantify the actual magnitude of $\Oasq$ effects. In
section~\ref{sec:discr} we provide a first indication that such effects are indeed small,
at least for the here considered light meson sector;
currently ongoing computations at a significantly smaller lattice spacing will allow for a 
continuum limit scaling analysis in this setup. 

\section{Simulation Details}
\label{sec:simdet}

\subsection{Simulation Ensembles}

We performed simulations at two values of the lattice gauge coupling
$\beta =1.90$ and $1.95$, corresponding to values of the lattice
spacing $a\approx 0.086$\,fm and $a\approx 0.078$\,fm, respectively.  The parameters
of each ensemble are reported in table~\ref{tab:par_sim}. The charged
pion mass $\mps$ ranges from $270$\,MeV to $510$\,MeV. 
Simulated volumes correspond to values of $\mps L$
ranging from $3.0$ to $5.8$, where the smaller volumes served to
estimate finite volume effects, see table~\ref{tab:tuning}. Physical spatial volumes 
range from $(1.9\,\mathrm{fm})^3$ to $(2.8\,\mathrm{fm})^3$. 

As already mentioned, the tuning to $\kappa_{crit}$ was performed
independently for each value of the mass parameters $a\mu_l$, $a\mu_\sigma$ and $a\mu_\delta$.
The mass parameters of the heavy
doublet $a\mu_\sigma$ and $a\mu_\delta$ reported in
table~\ref{tab:par_sim} are related to the strange and charm quark
masses. In particular, they are fixed by requiring the simulated kaon and $D$ meson
masses to approximately take their physical values, as discussed in
section~\ref{sec:KDmasses}.
\begin{table}[t!]
  \centering
  \begin{tabular*}{0.9\textwidth}{@{\extracolsep{\fill}}lcccccc}
    \hline\hline
    Ensemble & $\beta$ & $\kappa_\mathrm{crit}$ & $a\mu_l$ & $a\mu_\sigma$ & $a\mu_\delta$ & $(L/a)^3 \times T/a$\\
    \hline\hline
    A30.32&1.90&0.1632720&0.0030&0.150&0.190&$32^3 \times 64$\\
    A40.32&&0.1632700&0.0040&&&$32^3 \times 64$\\
    A40.24&&0.1632700&0.0040&&&$24^3 \times 48$\\
    A40.20&&0.1632700&0.0040&&&$20^3 \times 48$\\
    A50.32&&0.1632670&0.0050&&&$32^3 \times 64$\\
    A60.24&&0.1632650&0.0060&&&$24^3 \times 48$\\
    A80.24&&0.1632600&0.0080&&&$24^3 \times 48$\\
    A100.24&&0.1632550&0.0100&&&$24^3 \times 48$\\
    A100.24s&&0.1631960&0.0100&&0.197&$24^3 \times 48$\\
    \hline
    B25.32&1.95&0.1612420&0.0025&0.135&0.170&$32^3 \times 64$\\
    B35.32&&0.1612400&0.0035&&&$32^3 \times 64$\\
    B55.32&&0.1612360&0.0055&&&$32^3 \times 64$\\
    B75.32&&0.1612320&0.0075&&&$32^3 \times 64$\\
    B85.24&&0.1612312&0.0085&&&$24^3 \times 48$\\
    \hline
  \end{tabular*}
  \caption{Summary of the $N_{\rm f}=2+1+1$ ensembles generated by
    ETMC at two values of the lattice coupling $\beta =1.90$ and
    $\beta =1.95$. From left to right, we quote the ensemble name, the
    value of inverse coupling $\beta$, the estimate of the critical
    value $\kappa_{crit}$, the light twisted mass $a\mu_l$, the heavy
    doublet mass parameters $a\mu_\sigma$ and $a\mu_\delta$ and the
    volume in units of the lattice spacing. Our notation for the
    ensemble names corresponds to X.$\mu_l$.$L$, with X referring to the
    value of $\beta$ used. The run A100.24s is used to control the
    tuning of the strange and charm quark masses.}
  \label{tab:par_sim}
\end{table}
\begin{table}[t]
  \centering
  \begin{tabular*}{0.9\textwidth}{@{\extracolsep{\fill}}lccccc}
    \hline\hline
    Ensemble & $\mpcac/\mu_l$ & $\mps L$ & $\tauint(\langle P \rangle)$ & $\tauint(a\mps)$ & $\tauint(a\mpcac)$ \\
    \hline\hline
    A30.32& -0.123(87)& 3.97& 118(55) &2.7(4)&46(19)\\
    A40.32& -0.055(55)& 4.53& 103(48)&4.1(7)&51(21)\\
    A40.24& -0.148(83)& 3.48& 132(57)&$\leq$ 2&35(12)\\
    A40.20& -0.051(91)& 2.97&55(25) &2.9(7)&26(12)\\
    A50.32& 0.064(24) & 5.05& 50(19) &3.0(5)&21(7)\\
    A60.24& -0.037(50)& 4.15&28(8)&2.0(2)&13(4)\\
    A80.24& 0.020(19) & 4.77&23(7)&2.4(3)&10(2)\\
    A100.24& 0.025(18) & 5.35&18(5)&2.3(3)&13(3)\\
    A100.24s& 0.045(18) & 5.31&18(5)&6.2(1.1)&18(5)\\
    \hline
    B25.32& -0.185(69) & 3.42&65(25)&3.6(6)&26(9)\\
    B35.32& 0.009(34) & 4.03&54(19)&5.5(8)&41(14)\\
    B55.32& -0.069(13) & 4.97&12(3)&$\leq$ 2&8(2)\\
    B75.32& -0.047(12) & 5.77&14(4)&3.3(5)&13(3)\\
    B85.24& -0.001(16) & 4.66&15(4)&2.2(2)&11(2)\\
    \hline
  \end{tabular*}
  \caption{For each ensemble, from left to right the values of
    $\mpcac/\mu_l$, $\mps L$, the integrated autocorrelation time of
    the plaquette, $\mps$ and $\mpcac$ in units of the trajectory
    length. Every ensemble contains $5000$ thermalised trajectories of
    length $\tau=1$, except A40.24 which contains $8000$
    trajectories.}
  \label{tab:tuning}
\end{table}
The simulation algorithm used to generate the ensembles includes in
the light sector, a Hybrid Monte Carlo algorithm with multiple time
scales and mass preconditioning, described in
ref.~\cite{Urbach:2005ji}, while in the strange-charm sector a
polynomial hybrid Monte Carlo (PHMC)
algorithm~\cite{Frezzotti:1997ym,Frezzotti:1998eu,Chiarappa:2005mx};
the implementation of ref.~\cite{Jansen:2009xp} is publicly
available.

The positivity of the determinant of the Dirac operator is a
property of the mass-degenerate Wilson twisted mass action, which does not 
necessarily hold in the non degenerate case for generic values of the mass
parameters $\mu_\sigma$ and $\mu_\delta$.\,\footnote{Notice however that the positivity of
  the determinant is guaranteed for $\mu_\sigma^2 >
  \mu_\delta^2$~\cite{Frezzotti:2004wz,Frezzotti:2003xj}.} The
positivity is monitored by measuring the smallest
eigenvalue $\lambda_{\rm h,min}$ of $Q_h^\dagger Q_h$, where
$Q_h=\gamma_5 \tau_3 D_h$ and $D_h$ is the Wilson Dirac operator of
the non-degenerate twisted mass action in eq.~(\ref{eq:sf}). We
observe that $\lambda_{\rm h,min}$ is roughly proportional to the
renormalised strange quark mass squared. Since we choose the mass
parameters $\mu_\sigma$ and $\mu_\delta$ such that the strange quark
takes its physical value, a spectral gap in the distribution of
$Q_h^\dagger Q_h$ is observed, implying that the determinant of $D_h$
does not change sign during the simulation. 
While this is sufficient for the purpose of this study, we shall provide a detailed 
discussion of this issue in a forthcoming publication.

To generate correlators we use stochastic sources and improve the
signal-to-noise ratio by using the ``one-end trick'', following the
techniques also employed in our $N_{\rm f}=2$ simulations
\cite{Boucaud:2008xu}.  We have constructed all meson correlators with
local (L), fuzzed (F) and Gaussian smeared (S) sources and sinks.  The
use of smeared or fuzzed sources has stronger impact on the extraction
of the kaon and $D$ meson masses; results for the latter are reported
in section~\ref{sec:KDmasses}, while a companion
paper~\cite{Baron:HL2010} discusses the adopted strategy for the less
straightforward determination of these masses in the unitary $N_{\rm
  f}=2+1+1$ Wilson twisted mass formalism.

\subsection{Tuning to Maximal Twist}
\label{sec:tuning}

To guarantee $\Oa$ improvement of all physical observables while also
avoiding residual $\Oasq$ effects with decreasing pion mass, the
numerical precision of the tuning to maximal twist -- quantified by
the deviation from zero of $\mpcac$ -- has to satisfy $|Z_{\rm
  A}\mpcac/\mu_l|_{\mu_l,\,\mu_\sigma ,\,\mu_\delta} \lesssim a
\Lambda_{QCD}$~\cite{Boucaud:2007uk,Boucaud:2008xu,Baron:2009wt}. The
left-hand side contains the renormalised ratio of the untwisted mass
over the twisted light-quark mass. A similar condition should be
fulfilled by the error on this ratio. For the current lattice
spacings, $a\Lambda_{QCD} \approx 0.1$, while the values of the axial
current renormalisation factor $Z_{\rm A}$ have not yet been
determined. Nevertheless, since $Z_{\rm A}$ enters as an
$\mathcal{O}(1)$ multiplicative prefactor, and it is expected to be
$Z_{\rm A} \lesssim 1$ for our ensembles\footnote{Preliminary
  determinations of $Z_{\rm A}$ from ongoing dedicated runs with four
  degenerate light flavours, indicate that $Z_{\rm A}\sim 0.7 - 0.8$ for
  the ensembles considered in this work.}, we adopt the conservative
choice $Z_{\rm A}=1$ in verifying the tuning condition.
\begin{figure}[t]
  \includegraphics[width=0.92\linewidth]{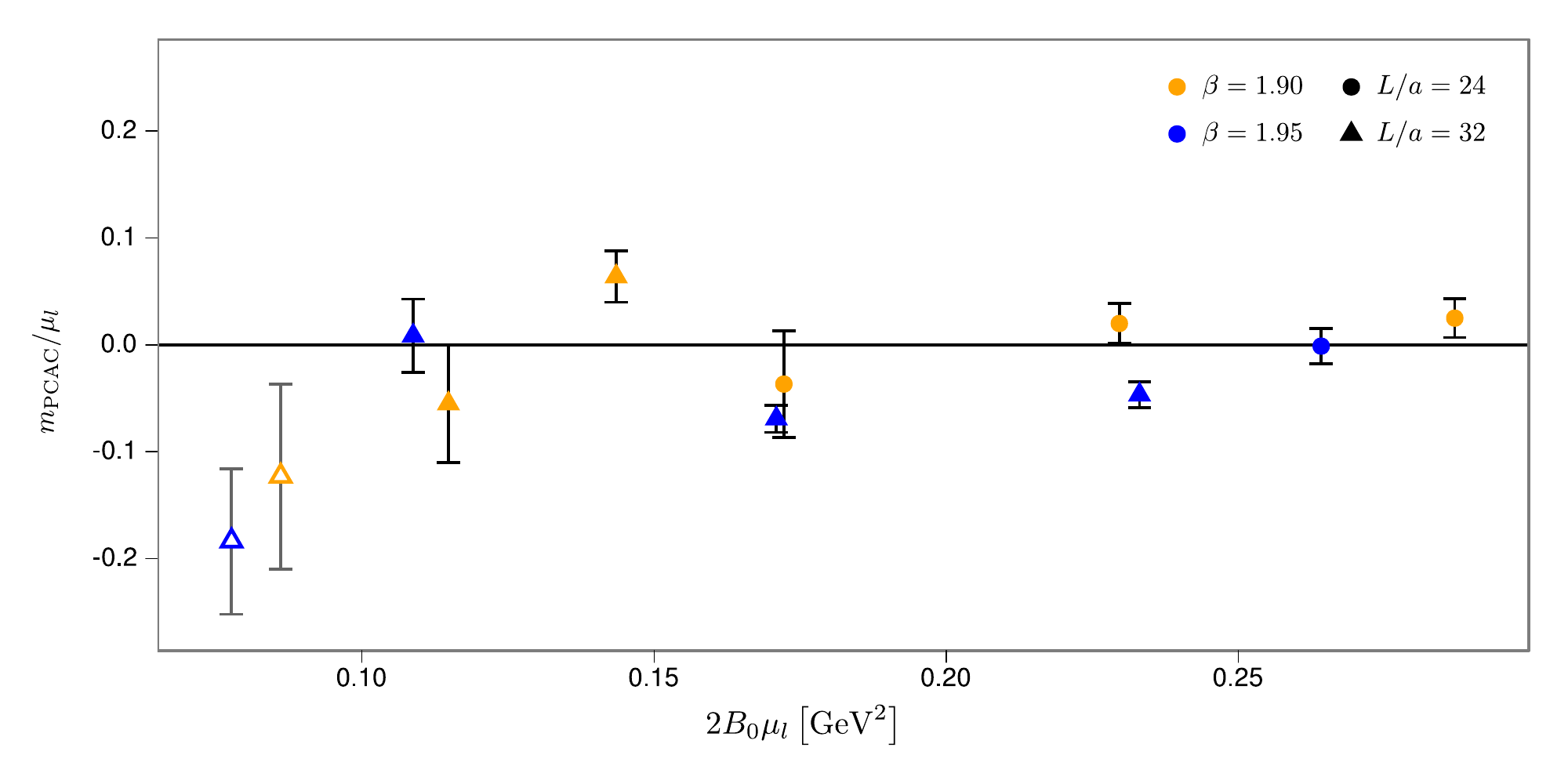}
  \caption{The ratio $\mpcac/\mu_l$ for the ensembles at $\beta=1.90$
    and $1.95$ at the largest simulated volumes and as a function of
    $2B_0\mu_l$. For both ensembles the ratio $\mpcac/\mu_l$ satisfies
    the $10\%$ level criterion, except for the lightest point at
    $\beta =1.90$ and $\beta =1.95$ (open symbols), also affected by
    larger statistical errors. We assume $Z_A =1$, while the actual
    value $Z_A\lesssim 1$ can only improve all tuning conditions.}
  \label{fig:PCACratiovsk}
\end{figure}

Satisfying this constraint clearly requires a good statistical
accuracy in the determination of the PCAC mass. The values of
$\mpcac/\mu_l$ reported in table~\ref{tab:tuning} and shown in
fig.~\ref{fig:PCACratiovsk} are well satisfying the tuning condition
to maximal twist, with the exception of the lightest mass point at
$\beta =1.90$ and $\beta =1.95$. We notice that the autocorrelation
time of $\mpcac$ reported in table~\ref{tab:tuning} grows with
decreasing values of the light quark mass $\mu_l$, thus rendering the
tuning more costly for the two lightest points.  For the ensemble
B25.32, we are currently performing a new simulation aiming at a more
accurate tuning to $\kappa_{\rm crit}$. We are also testing a
reweighting procedure~\cite{Baron:2008xa} in $\kappa$ on the same ensemble, in view of
applying it to the other not optimally tuned ensemble A30.32, and to
future simulations.
In what follows, we use the lightest mass points for consistency checks, and
we exclude them from the final $\chi$PT fits.  We also remind the
reader that the small deviations from zero of $a\mpcac$ will only
affect the $\Oasq$ lattice discretisation errors of physical
observables~\cite{Boucaud:2008xu}.
\subsection{Tuning of the Strange and Charm Quark Masses}
\label{sec:KDmasses}

The mass parameters $\mu_\sigma$ and $\mu_\delta$ in the heavy doublet
of the action in eq.~(\ref{eq:sf}) can in principle be adjusted so as
to match the renormalised strange and charm quark masses by use of
eq.~(\ref{eq:msmc}). In practise, in this work, we fix the values of
$\mu_\sigma$ and $\mu_\delta$ by requiring that the simulated kaon
mass $m_K$ and $D$ meson mass $m_D$ approximately take their physical
values.

\begin{figure}[t]
\centering
  \subfigure[\label{fig:mk}]
  {\includegraphics[width=0.47\linewidth]{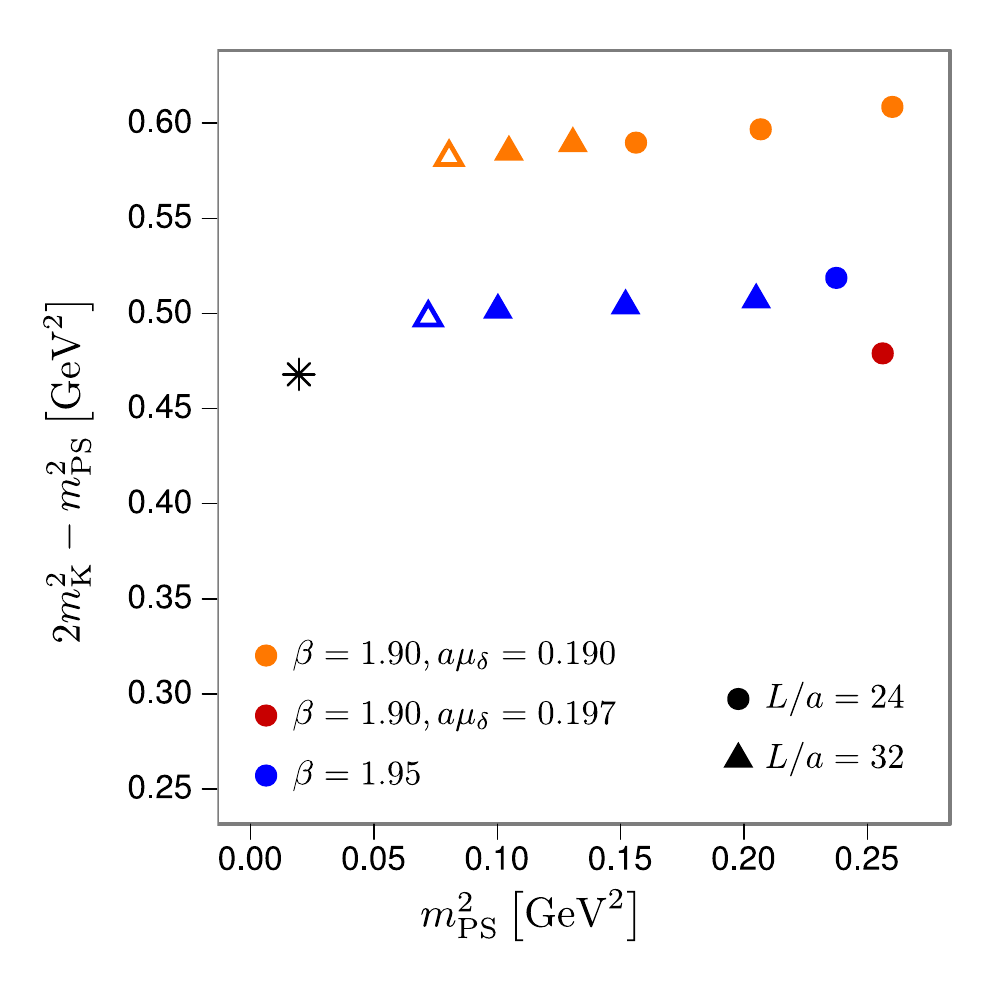}}
  \subfigure[\label{fig:mD}]
  {\includegraphics[width=0.47\linewidth]{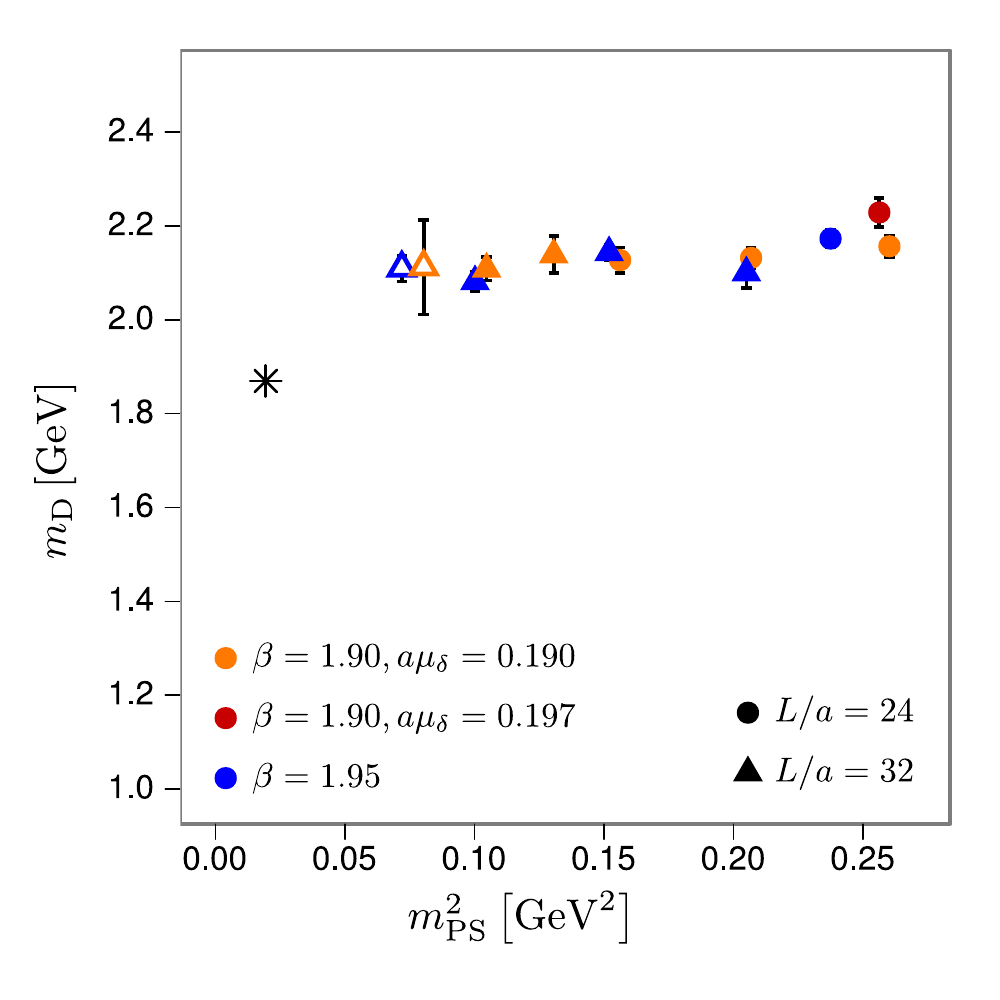}}
  \caption{(a): $2m_K^2 -\mps^2$, and
    (b): $m_D$, as a function of $\mps^2$, for $\beta=1.95$ (blue) and
    $\beta=1.90$ (orange). The physical point is shown (black star).
    The kaon and $D$ meson masses appear to be properly tuned at
      $\beta=1.95$. The ensembles at $\beta=1.90$,
      $\mu_\delta=0.190$ have a larger value of the strange
      quark mass, while the red point at $\beta =1.90$, $a\mu_\delta =
      0.197$ appears to be well tuned.
    Data points have been scaled with the
    lattice spacing $a=0.08585(53)$\,fm for $\beta=1.90$, and
    $a=0.07820(59)$\,fm for $\beta=1.95$, obtained in this work and where the errors are
      only statistical. }
\end{figure}

A detailed description of the determination of the kaon and $D$ meson
masses is separately given in~\cite{Baron:HL2010}, while
figures~\ref{fig:mk}~and~\ref{fig:mD} show the resulting dependence of
$(2 m_K^2-m_{PS}^2)$ and $m_D$ upon the light pseudoscalar mass
squared for both ensembles, and compared with the physical point.
Table~\ref{tab:kdmasses} summarises their numerical values, while the
corresponding values for $a\mu_\sigma$ and $a\mu_\delta$ are given in
table~\ref{tab:par_sim}.  Observe also that, in order to be able to
properly tune the strange and charm quark masses to their physical
values, $a\mu_\sigma$ must be chosen larger than $a\mu_\delta$, since
(see eq. (6)) the ratio $Z_P/Z_S$ is significantly smaller than
one~\cite{Baron:HL2010}.

\begin{table}
  \centering
  \begin{tabular*}{0.7\textwidth}{@{\extracolsep{\fill}}lccc}
    \hline\hline
    Ensemble & $\beta$ & $am_K$ & $am_D$ \\
    \hline\hline
    A30.32& 1.90 &0.25150(29)&0.9230(440)\\
    A40.32&&0.25666(23)&0.9216(109)\\
    A40.24&&0.25884(43)&0.9375(128)\\
    A40.20&&0.26130(135)&0.8701(152)\\
    A50.32&&0.26225(38)&0.9348(173)\\
    A60.24&&0.26695(52)&0.9298(118)\\
    A80.24&&0.27706(61)&0.9319(94)\\
    A100.24&&0.28807(34)&0.9427(99)\\
    A100.24s&&0.26502(90)&0.9742(133)\\
    \hline
    B25.32& 1.95&0.21240(50)& 0.8395(109)\\
    B35.32&&0.21840(28)& 0.8286(85)\\
    B55.32&&0.22799(34)& 0.8532(62)\\
    B75.32&&0.23753(32)& 0.8361(127)\\
    B85.24&&0.24476(44)& 0.8650(76)\\
    \hline
  \end{tabular*}
  \caption{For each ensemble, the values of the kaon mass 
    and the $D$ meson mass as determined in~\cite{Baron:HL2010}.}
  \label{tab:kdmasses}
\end{table}
While the kaon and $D$ meson masses at $\beta=1.95$ are sufficiently
well tuned to their physical values, the ensembles at $\beta=1.90$
with $a\mu_\delta=0.190$ carry a heavier kaon mass. The latter is
instead visibly closer to its physical value for $a\mu_\delta=0.197$,
as can be inferred from figure \ref{fig:mk}. We are currently
performing simulations with $a\mu_\delta=0.197$ for other light quark
masses. Moreover, another set of values of $\mu_\sigma$ and
$\mu_\delta$ are currently being used at $\beta=1.90$ to generate
ensembles with a slightly lower $D$ meson mass and a third value of
the kaon mass, in order to properly interpolate the lattice data to
the physical strange quark mass.
\subsection{Discretisation Effects in Light-quark Observables}
\label{sec:discr}

 In this section we explore discretisation effects in the analysed
 light-quark observables. To this aim we also make use of the
 determination of the chirally extrapolated $r_0$ value for our
 data samples, as discussed in the following section~\ref{sec:r0}.

\begin{figure}
\centering
  \subfigure[\label{fig:discr}]
  {\includegraphics[width=0.47\linewidth]{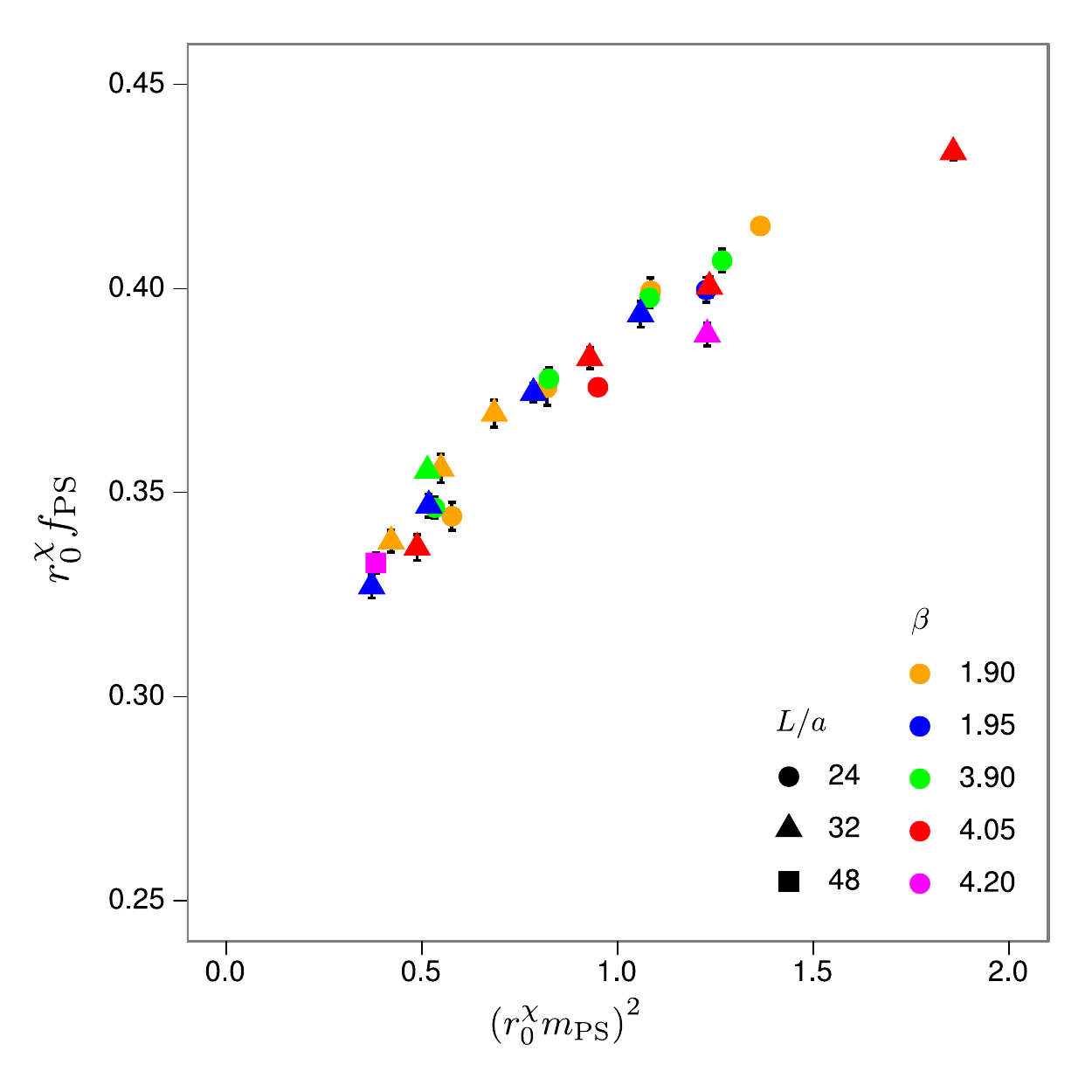}}
  \subfigure[\label{fig:discr2}]
  {\includegraphics[width=0.47\linewidth]{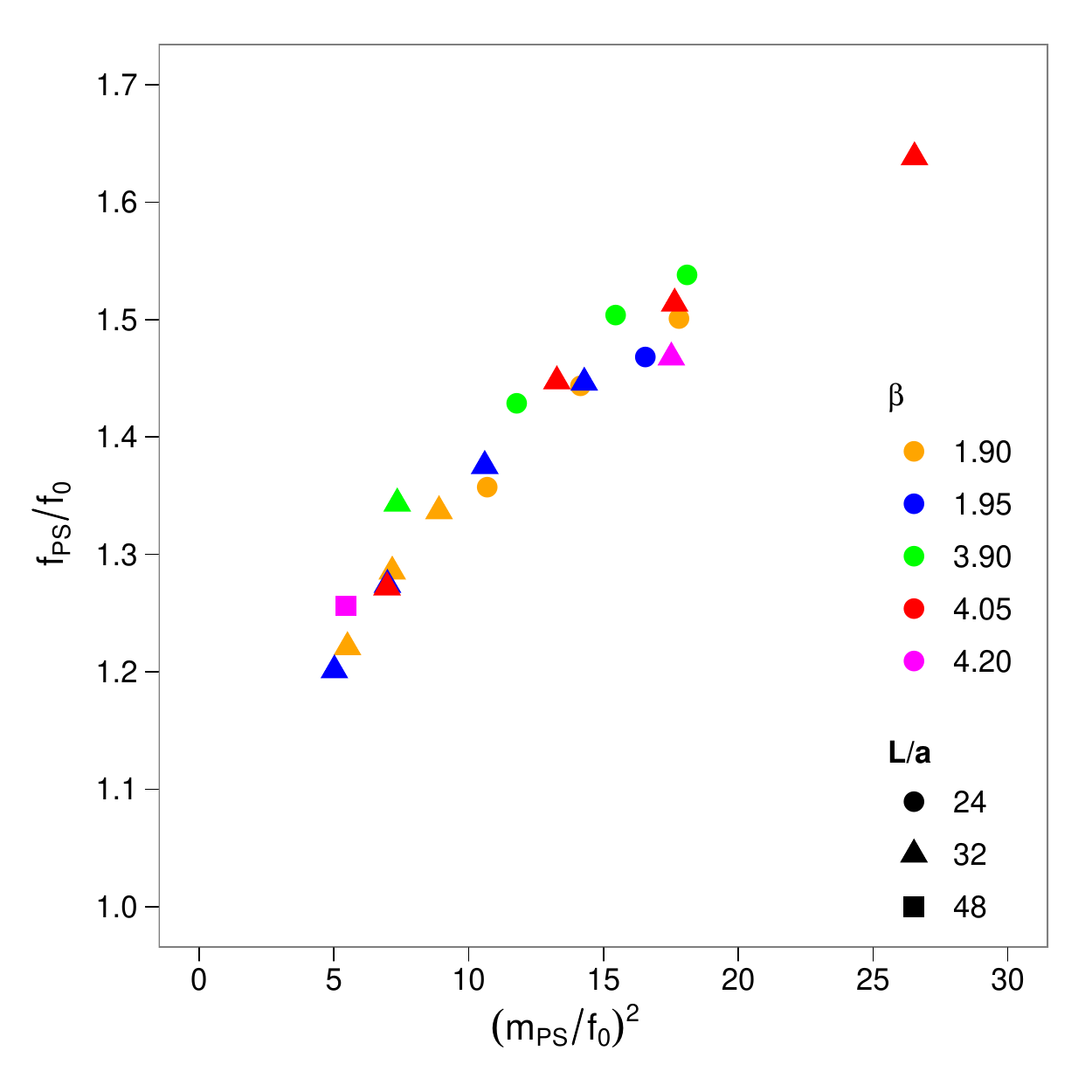}}
  \caption{
The quantity  $\alpha \fps$ as a function of $(\alpha\mps)^2$, 
with (a) $\alpha = r_0^\chi$ and (b) $\alpha = 1/f_0$,
for the $N_{\rm f}=2+1+1$ data at $\beta =1.90$ and 
    $\beta =1.95$, and for the $N_{\rm f} =2$ data at $\beta=3.90$,
    $\beta=4.05$ and $\beta=4.20$ in~\cite{Baron:2009wt}. 
The values of $r_0^\chi$ for $N_f=2+1+1$ are given in
      tables~\ref{table:fitresults} and \ref{table:fitresults_c}.
}
\end{figure}
\begin{figure}
  \centering
  \includegraphics[width=0.50\linewidth]{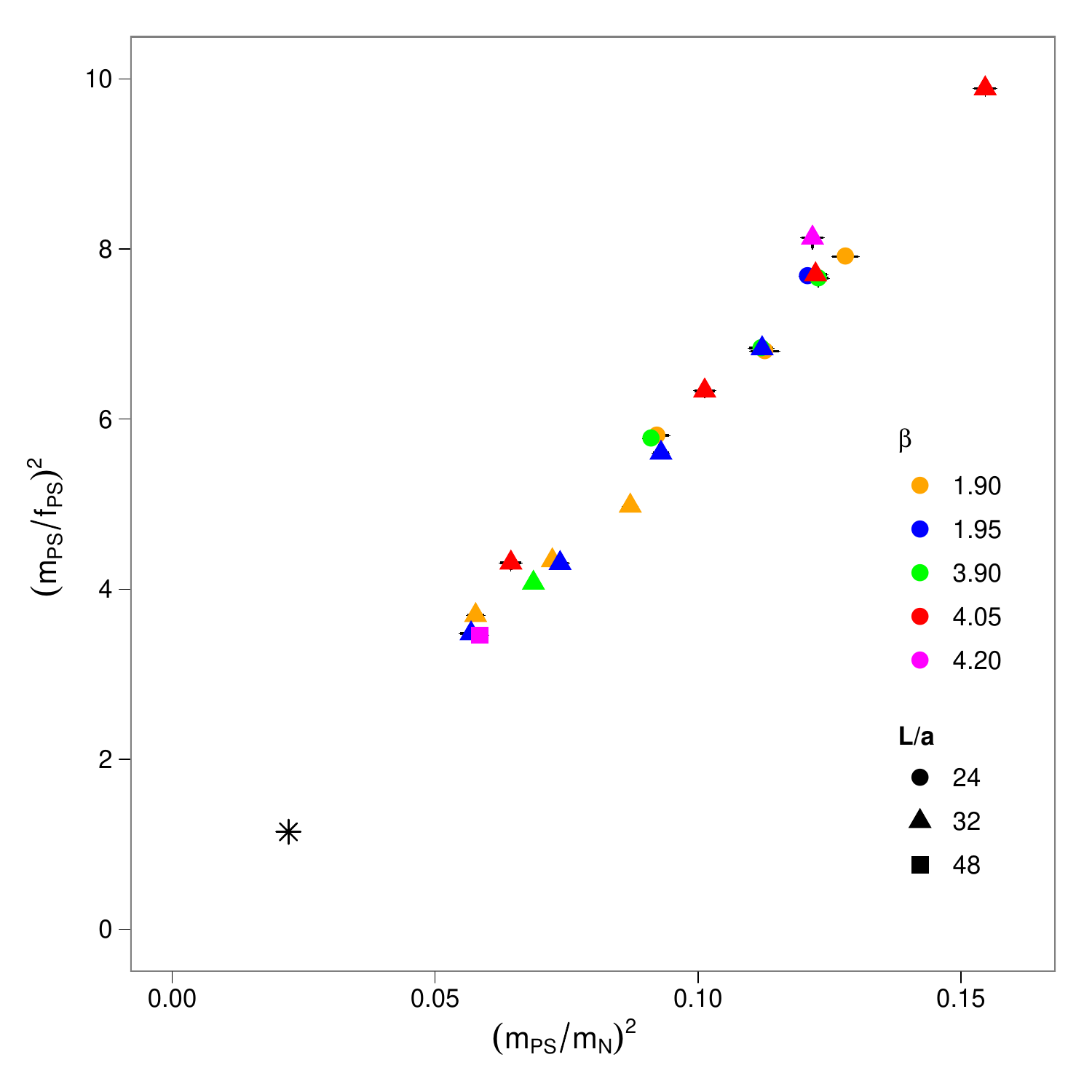}
  \caption{The ratio $\mps^2/\fps^2$ as a function of $\mps^2/m_N^2$,
    for the $N_{\rm f}=2+1+1$ ensembles at $\beta =1.90$ and $\beta
    =1.95$, compared to the $N_{\rm f} =2$ data at $\beta=3.90$,
    $\beta=4.05$ and $\beta=4.20$~\cite{Baron:2009wt}. The physical
    point is shown (black star).}
  \label{fig:nuclratio}
\end{figure}

In figures~\ref{fig:discr} and \ref{fig:discr2} we study the
sensitivity of the charged pion mass and decay constant to possible
discretisation effects, by comparing the $N_{\rm f}=2+1+1$ data at
$\beta=1.90$ and $\beta=1.95$ and the results obtained in twisted mass
simulations with two dynamical flavours~\cite{Baron:2009wt}. The
alignment of all data points at different values of $\beta$ is in
itself an indication of small discretisation effects.  The comparison
and good agreement with the $N_{\rm f}=2$ data seems also to suggest
no significant dependence upon the inclusion of dynamical strange and
charm quarks for these light observables, at least at the present
level of accuracy and provided that no cancellations occur due to
lattice discretisation effects. However, only a more complete study at
significantly different lattice spacings will allow to draw
conclusions.

In the same spirit, we show in figure~\ref{fig:nuclratio} an analogous
ratio plot where the nucleon mass data points are included. The
alignment of all data and the good extrapolation to the physical point
is again evident.
We defer to future publications the analysis of the baryon
spectrum and the study of discretisation effects in strange- and
charm-quark observables.

\subsection{The Sommer Scale $r_0$}
\label{sec:r0}

The Sommer
  scale $r_0$~\cite{Sommer:1993ce} is a purely gluonic quantity
  extracted from the static inter-quark potential. Since the knowledge
  of its physical value remains rather imprecise, we use the chirally
  extrapolated lattice data for $r_0/a$ only as an effective way to
  compare results from different values of the lattice spacing. In
  this work, the lattice scale is extracted by performing $\chi$PT
  inspired fits to the very precise data for $a\fps$ and $a\mps$, and
  by using the physical values of $m_\pi$ and $f_\pi$ as inputs.
\begin{figure}[t]
  \centering
   \subfigure[\label{fig:r0overa}]
  {\includegraphics[width=0.47\linewidth]{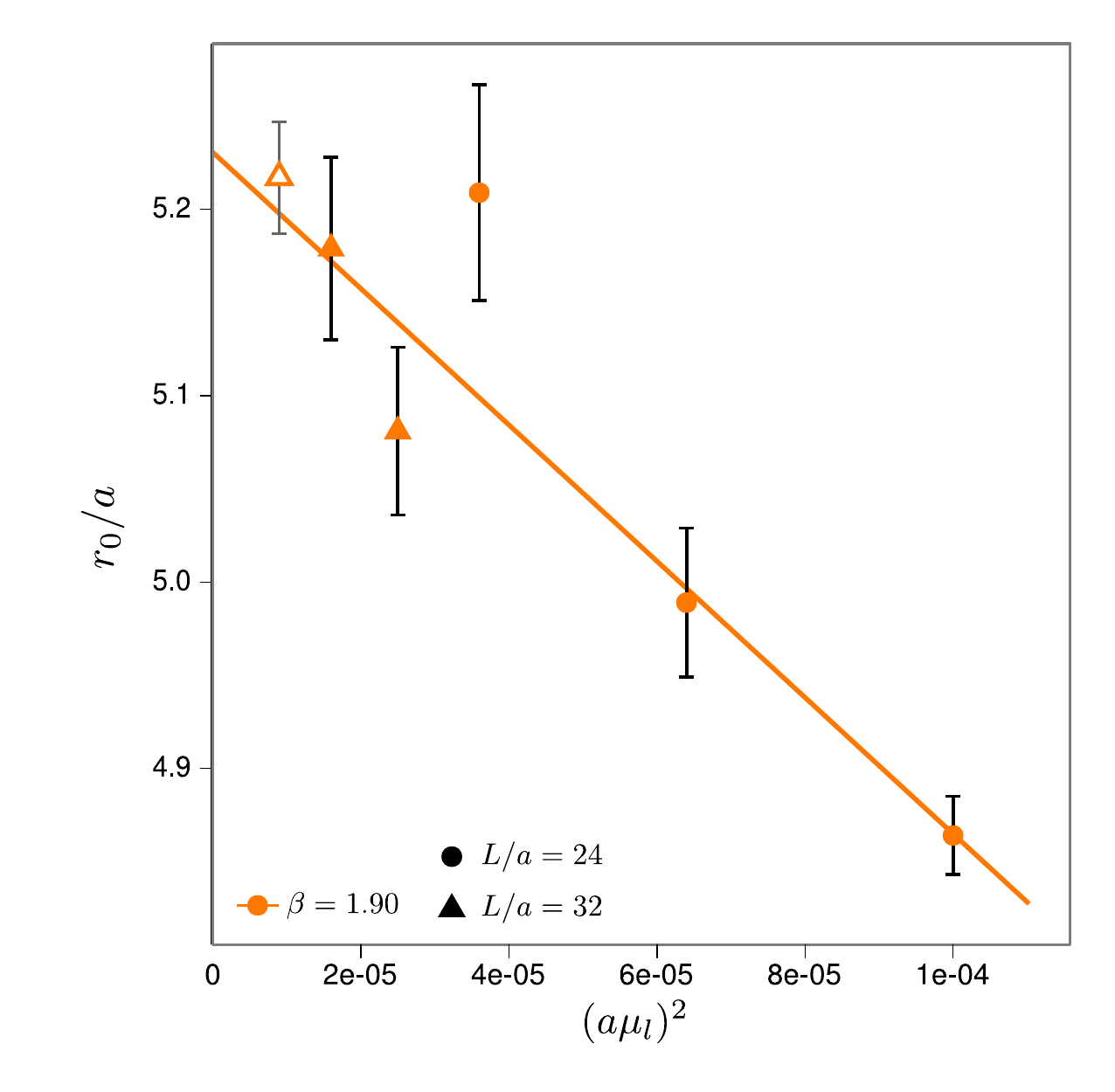}}
  \subfigure[\label{fig:r0overa_2}]
  {\includegraphics[width=0.47\linewidth]{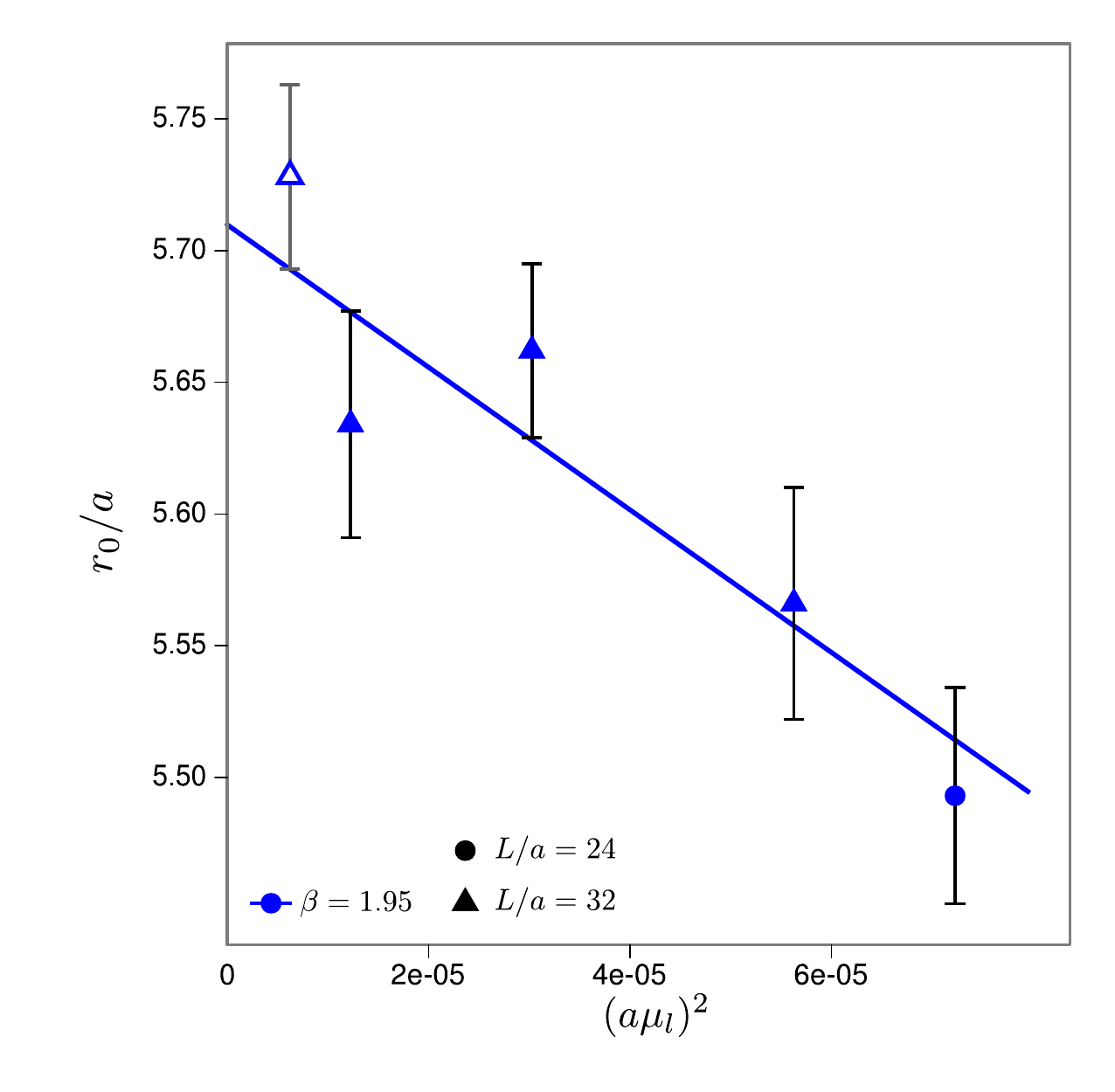}} 
  \caption{The Sommer scale $r_0/a$ as a function of $(a\mu_l )^2$
    for (a) $\beta=1.90$ and (b) $\beta =1.95$.  The lines
    represent a linear extrapolation in $(a\mu_l )^2$ to the chiral
    limit. The lightest point (open symbol) is not included in the fits 
and we have always used the largest available volume
    for a given value of the mass.}
\end{figure}

Figures~\ref{fig:r0overa} and \ref{fig:r0overa_2} display the data for
$r_0/a$ at both values of the lattice coupling $\beta =1.90$ and
$1.95$, and as a function of the bare lattice mass squared. The data
are reasonably well described by a quadratic dependence, as also
previously found for our $N_{\rm f}=2$ ensembles.  For a more detailed
discussion of the possible functional forms and their theoretical
interpretation see~\cite{Baron:2009zq}.  To extrapolate to the chiral
limit, we
have performed fits using the largest available volume at each value
of the pseudoscalar mass.  The chirally extrapolated values for our
$N_{\rm f}=2+1+1$ ensembles are $r_0^\chi /a = 5.231(38)$ at $\beta
=1.90$ and $r_0^\chi /a = 5.710(41)$ at $\beta =1.95$, where the
lightest points of both ensembles have been excluded from the
extrapolation, consistently with the fact that they do not satisfy our
most stringent tuning condition to maximal twist.

In order to meaningfully compare the dependence upon the light quark mass at the 
two different lattice couplings $\beta =1.90$ and $1.95$, we estimated the slope of the functional form 
$r_0/r_0^\chi =1 + c_r (r^\chi_0\mps )^4$, where the explicit lattice spacing dependence 
has been removed. We observe a mild
dependence on the light quark mass and similar slopes 
 $c_r[\beta =1.90 ]= -0.0379(37)$ and $c_r[\beta =1.95 ]=-0.0234(69)$.
It is also worth noticing that the dependence upon the light quark mass of the 
$N_{\rm f}=2+1+1$ data and that observed in the 
$N_{\rm f}=2$ case \cite{Baron:2009zq} are not significantly different.

\subsection{Effects of Isospin Breaking}\label{subsec:isospin}

A most delicate aspect of the twisted mass formulation is the breaking
of the isospin symmetry. Clear evidence for this breaking has been
found in the $N_{\rm f} = 2$ simulations by ETMC when comparing the
neutral with the charged pion masses. Indeed, while the discretisation
effects in the charged pion were observed to be very small,
significant $\mathcal{O}(a^2)$ corrections appear when studying the
scaling to the continuum limit of the neutral
pion~\cite{Baron:2009wt}. Notice, however, that similar effects have
not been observed in other quantities that are in principle sensitive
to isospin breaking but not trivially related to the neutral pion
mass. These observations are supported by theoretical considerations
detailed in~\cite{Frezzotti:2007qv,Dimopoulos:2009qv}.

In the $N_{\rm f}=2+1+1$ case, it turns out that the isospin breaking
effect in the mass difference of charged and neutral pion masses is
larger than for $N_{\rm f}=2$ at fixed physical
situation\footnote{Notice however that different gauge actions are
  used in the $N_{\rm f}=2$ and $N_{\rm f}=2+1+1$ cases as described
  in section~\ref{sec:gauge}.}, as can be inferred from
table~\ref{tab:neutral}.  On the other hand, the same theoretical
considerations as in~\cite{Dimopoulos:2009qv} do apply to the case of
$N_{\rm f}=2+1+1$ flavours, and it is expected that the same
class of physical observables as for $N_f=2$ will not be significantly
affected by isospin breaking corrections.  Having said that, a careful
measure of this effect for each observable or class of
observables is anyway mandatory.
\begin{table}[t]
  \centering
  \begin{tabular*}{0.9\textwidth}{@{\extracolsep{\fill}}lcccc}
    \hline\hline
    Ensemble & $\beta$ & $r^\chi_0\mps^{\pm}$ & $r^\chi_0m_{\mathrm{PS}}^0$ & $c$ \\
    \hline\hline
    B35.32 & 1.95 &   0.7196(57)   & 0.388(40)  &  -12.0(1.1)   \\
    B55.32 &      &   0.8861(67) &   0.679(40)  &  -10.6(1.8) \\
    \hline
    $B_6$ $N_f=2$ & 3.90  & 0.7113(66)  & 0.585(43)  & -4.6(1.5) \\
    $B_2$ $N_f=2$ &  & 0.9001(86) & 0.712(54) &  -8.6(2.2)\\
    \hline
  \end{tabular*}
  \caption{Measurements of the masses of the charged and the neutral
    pion. We compare runs at $\beta =1.95$ and $N_f =2$
    runs~\cite{Baron:2009wt} with comparable lattice spacing and
    similar charged pion masses in physical units. All masses are
    reported in units of the chirally extrapolated $r_0$ for the same
    ensemble, see table~\ref{table:fitresults_c}, and $r_0^\chi /a =
    5.316(49)$ for $N_f=2$. We also report on the approximate value of
    $c$, giving the slope of the $a^2$ dependence of the pion mass
    splitting.}
  \label{tab:neutral}
\end{table}
The increase of the pion mass splitting with increasing the number of
flavours in the sea is in line with the
observation~\cite{Chiarappa:2006ae} of a stronger first order phase
transition when moving from $N_{\rm f}=2$ to $N_{\rm f}=2+1+1$, as
discussed in section~\ref{sec:gauge}. Indeed, the endpoint of the phase
transition~\cite{Munster:2003ba,Sharpe:2004ny} corresponds to the
critical value of the light twisted mass $\mu_{l,{\rm c}}$ where the
neutral pion mass vanishes. The mass difference can be described by
$r_0^{\chi 2}((\mps^0)^2 - (\mps^\pm)^2 ) = c\, (a/r^\chi_0)^2$, 
where the coefficient $c$ is related to
$\mu_{l,{\rm c}}$~\cite{Munster:2003ba,Sharpe:2004ny} 
and it is therefore a measure of the strength of the
first order phase transition.  Hence, a larger value of $c$ means that
simulations are to be performed at smaller values of the lattice
spacing to reach, say, the physical point.
Table
\ref{tab:neutral} reports on the values of $\mps^\pm$, $\mps^0$ and
$c$ for some examples taken from the $\beta =1.95$ ensemble and the
$N_f=2$ ensemble with the closest values of the lattice spacing and physical 
charged pseudoscalar mass. As anticipated, the coefficient $c$ increases in 
absolute value from $N_{\rm f}=2$ to $N_{\rm f}=2+1+1$.

We are currently performing simulations at a significantly different
and lower lattice spacing than the present ensembles.  They will allow
to determine the slope $c$ for $N_{\rm f} = 2+1+1$ more accurately and
to better quantify the conditions to approach the physical point.

\subsection{Stout Smeared Runs}
\label{sec:stout}

In addition to our main simulation ensembles, we also performed runs
with stout smeared gauge fields in the lattice fermionic action.  The
stout smearing as introduced in~\cite{Morningstar:2003gk} was designed
to have a smearing procedure which is analytic in the unsmeared link
variables and hence well suited for HMC-type updating algorithms.  In
an earlier work with $N_{\rm f}=2$ quark flavours~\cite{Jansen:2007sr}
we showed that using smeared gauge fields in the fermion operator is
reducing the strength of the phase transition in twisted quark mass
simulations and therefore allows to reach smaller quark masses at a
given lattice spacing.

The definition of the stout smeared links can be found
in~\cite{Morningstar:2003gk}, and for the parameter $\rho$ connecting
thin to fat gauge links we choose $\rho =0.15$.  In principle, such
smearing can be iterated several times, with the price of rendering
the fermion action delocalised over a larger lattice region. We made a
conservative choice to maintain the action well localised and
performed a single smearing step.  As shown in~\cite{Jansen:2007sr},
this kind of smearing does not substantially change the lattice
spacing, and for the sake of comparison we thus kept the same value of
$\beta$ as in one of the non stout-smeared runs.  On the other hand,
the hopping parameter has to be tuned again, since the additive
renormalisation of the quark mass is expected to be smaller.  The
parameters of our runs are given in Table~\ref{tab:stoutruns}.  These
runs have been done with the two-step polynomial Hybrid Monte Carlo
(TS-PHMC) update algorithm~\cite{Montvay:2005tj}.
\begin{table}[t!]
  \centering
  \begin{tabular*}{0.9\textwidth}{@{\extracolsep{\fill}}lccccccc}
    \hline\hline
    Ensemble & $\beta$ & $\kappa_{\rm crit}$ & $a\mu_l$ &
    $a\mu_\sigma$ & $a\mu_\delta$ & $N_{\rm traj.}$ & $r_0/a$
    \\ \hline\hline
    A$_{\rm st}$40.24 & 1.90 & 0.145512 & 0.0040 & 0.170 & 0.185 &
    { 1500} & 5.304(35)
    \\ 
    A$_{\rm st}$60.24 && 0.145511 & 0.0060 &&& { 3100} & 5.300(37)
    \\ 
    A$_{\rm st}$80.24 && 0.145510 & 0.0080 &&& { 2000} & 5.353(43)
    \\\hline
  \end{tabular*}
  \caption{Parameters of the runs with stout smearing on
    $L/a=24$, $T/a=48$ lattices. The number of thermalised trajectories with
    length $\tau =1$ is given by $N_{\rm traj.}$. The label ``{\rm st}'' in the
    ensemble name refers to the use of stout smearing, compared to the non stout-smeared 
ensemble in table~\ref{tab:par_sim}.}
  \label{tab:stoutruns}
\end{table}
\begin{table}
  \centering
  \begin{tabular*}{0.9\textwidth}{@{\extracolsep{\fill}}lcccc}
    \hline\hline
    Ensemble & $am_{\rm PS}$ &$am_K$ &$am_D$ &$\mpcac/\mu_l$
    \\ \hline\hline
    A$_{\rm st}$40.24& 0.12600(93) & 0.2479(18)  & 0.802(27) & 0.0175(68)    \\ 
    A$_{\rm st}$60.24& 0.14888(78) & 0.25338(67) & 0.825(26) & 0.0017(50)    \\ 
    A$_{\rm st}$80.24& 0.17156(69) & 0.26198(80) & 0.811(12) & 0.0138(48)
    \\\hline
  \end{tabular*}
  \caption{The masses in lattice units for the ensembles with one level of stout 
    smearing.}
  \label{tab:stoutmasses}
\end{table}
Results for the hadron masses are collected in
Table~\ref{tab:stoutmasses}, where the quoted errors include an
estimate of the systematic error induced by variations of the fitting
range.  The method of estimating and combining statistical and
systematic errors for the case of the kaon and $D$ meson masses is
described in~\cite{Baron:HL2010}.

As the values of $\mpcac/\mu_l$ in table~\ref{tab:stoutmasses} show,
the hopping parameters are well tuned to maximal twist.  The masses in
the run with smallest light twisted mass $a\mu_l=0.0040$ (ensemble
A$_{\rm st}$40.24) satisfy $r_0 \mps = 0.668(10)$, $r_0 m_K =
1.315(13)$ and $r_0 m_D = 4.25(29)$.  This means that the pion is
lighter than in the corresponding run without stout smearing (see
table~\ref{tab:rawdata}) and the kaon and $D$ meson masses are closer
to their physical value.  The smaller pion mass should be interpreted
as due to a quark mass renormalisation factor closer to one.  For the
same reason the tuned twisted masses in the heavy doublet $a\mu_\sigma
= 0.170$, $a\mu_\delta=0.185$ are smaller than in the runs without
stout smearing.  It is also interesting to compare the mass splitting
of the charged and neutral pion between runs with and without stout
smearing.  For the ensemble A$_{\rm st}$60.24 we obtain a neutral pion
mass $r^\chi_0\mps^0 = 0.409(34)$ and a charged pion mass
$r^\chi_0\mps^\pm = 0.7861(56)$, in units of the chirally extrapolated
value $r_0^\chi/a =5.280(25)$, providing an estimate of the slope $c =
-12.6(0.8)$.  Notice that the mass dependence of $r_0/a$ in
table~\ref{tab:stoutruns} is reduced as compared to the runs with no
stout smearing, and a quadratic dependence on the bare quark mass has
been used for the extrapolation to the chiral limit, consistently with
the analysis of section \ref{sec:r0}.  For the corresponding ensemble
A60.24 without stout smearing, using data in tables~\ref{tab:rawdata}
and \ref{table:fitresults_c}, we obtain instead $r^\chi_0\mps^0 =
0.560(37)$, $r^\chi_0\mps^\pm = 0.9036(71)$, and a slope $c =
-13.8(1.2)$, slightly but not significantly different from the
stout-smeared case.

The runs with stout-smeared gauge links show somewhat better characteristics
than the ones without stout smearing, but the improvements are not dramatic, at least 
with one level of stout smearing. More iterations would further accelerate 
the approach to lighter masses and are expected to further reduce the charged to 
neutral pion splitting. However, it is a delicate matter to establish how physical 
observables other than the spectrum will be affected. 
Based on these considerations and given the present pool of data, the 
final results in this study are obtained with non stout-smeared 
simulations. 

\section{Results: {\bf $\fps$},  {\bf $\mps$} and Chiral Fits}
\label{sec:observables}

We concentrate in this section on the analysis of the simplest and
phenomenologically relevant observables involving up and down
valence quarks. These are the light charged pseudoscalar decay constant $\fps$
and the light charged pseudoscalar mass $\mps$. 

The present simulations with dynamical strange and charm quarks, sitting
at, or varying around, their nature given masses, should allow for a good measure of the
impact of strange and charm dynamics on the low energy sector of
QCD and the electroweak matrix elements. As a first step, one can determine 
the low energy constants of chiral perturbation theory ($\chi$PT).  
The values of $a\fps$ and $a\mps$ for our ensembles at $\beta =1.95$ and $\beta =1.90$ are
summarised in table~\ref{tab:rawdata}.
\begin{table}[t!]
  \centering
  \begin{tabular*}{0.9\textwidth}{@{\extracolsep{\fill}}lccccc}
    \hline\hline
    Ensemble & $a\mu_l$ & $am_\mathrm{PS}$ & $af_\mathrm{PS}$ &
    $r_0/a$ & $L/a$ \\
    \hline\hline
    A30.32  &0.0030&0.12395(36)(14)&0.06451(35)(3)&5.217(30)&32\\
    A40.32  &0.0040&0.14142(27)(42)&0.06791(18)(4)&5.179(49)&32\\
    A40.24  &0.0040&0.14492(52)(34)&0.06568(34)(7)&5.178(44)&24\\
    A40.20  &0.0040& 0.14871(92)(116) & 0.06194(65)(23)&-&20\\
    A50.32  &0.0050&0.15796(32)(28)&0.07048(16)(4)&5.081(45)&32\\
    A60.24  &0.0060&0.17275(45)(23)&0.07169(22)(2)&5.209(58)&24\\
    A80.24  &0.0080&0.19875(41)(35)&0.07623(21)(4)&4.989(40)&24\\
    A100.24 &0.0100&0.22293(35)(38)&0.07926(20)(4)&4.864(21)&24\\
    A100.24s&0.0100&0.22125(58)(119)&0.07843(26)(21)&4.918(50)&24\\
    \hline
    B25.32  &0.0025&0.10680(39)(27)&0.05727(36)(8)&5.728(35)&32\\
    B35.32  &0.0035&0.12602(30)(30)&0.06074(18)(8)&5.634(43)&32\\
    B55.32  &0.0055&0.15518(21)(33)&0.06557(15)(5)&5.662(33)&32\\
    B75.32  &0.0075&0.18020(27)(3)&0.06895(17)(1)&5.566(44)&32\\
    B85.24  &0.0085&0.19396(38)(54)&0.06999(20)(5)&5.493(41)&24\\
    \hline
  \end{tabular*}
  \caption{ Lattice measurements of the charged pseudoscalar mass
    $a\mps$, the charged pseudoscalar decay constant $a\fps$ and the
    Sommer scale in lattice units $r_0/a$ for our two ensembles at
    $\beta =1.90$ (A set) and $\beta =1.95$ (B set). The value of the
     light twisted mass $a\mu_l$ and the spatial length $L/a$ are also
    shown. Quoted errors are given as (statistical)(systematic), with the 
estimate of the systematic error coming from the uncertainty related to the 
fitting range.}
  \label{tab:rawdata}
\end{table}
In contrast to standard Wilson fermions, an exact lattice Ward
identity for maximally twisted mass fermions allows for extracting the
charged pseudoscalar decay constant $\fps$ from the relation
\begin{equation}
  \fps = \frac{2\mu_l}{\mps^2} |\langle 0| P_l^1 (0)|\pi\rangle |\,,
\end{equation}
without need to specify any renormalisation factor, since $Z_{\rm P} =
1/Z_{\mu}$~\cite{Frezzotti:2000nk}. We have performed fits to NLO
SU(2) continuum $\chi$PT at $\beta =1.95$ and $\beta =1.90$,
separately and combined. Results are summarised in table~\ref{table:fitresults_c}.

We thus simultaneously fit our data for the pseudoscalar mass and
decay constant to the following formulae, where the contributions $F$,
$D$ and $T$ parametrising finite size corrections, discretisation
effects and NNLO $\chi$PT effects, respectively, will be discussed
below:
\begin{eqnarray}
\mps^2 (L) &=& \chi_\mu\left(1 +    \xi\,l_3 + D_{\mps^2} a^2 + \xi^2\,T_{\mps^2} \right) 
F_{\mps^2} \nonumber \\
\fps (L) &=&   f_0     \left(1 - 2\,\xi\,l_4 + D_{\fps} a^2 + \xi^2\,T_{\fps} \right) 
F_{\fps}, \label{eq:mpfp}
\end{eqnarray}
with the pseudoscalar mass squared at tree level defined as $\chi_\mu
\equiv 2\,B_0\,\mu_l$ and the chiral expansion parameter by $\xi
\equiv \chi_\mu/\left(4\pi f_0\right)^2$. The low energy constants
$l_3$ and $l_4$ receive renormalization corrections
according to $\bar{l}_i = l_i + \ln\left[{\Lambda^2}/{\chi_\mu}\right]$, 
with $\Lambda$ the reference scale. During the fitting procedure, where all quantities
are defined in lattice units, we set the reference scale to a single
lattice spacing to let its constant logarithmic contribution
vanish. Once the scale of the simulation has been set, the low energy
constants are rescaled to the scale of the physical pion mass to
recover the physical values $\bar{l}_3$ and $\bar{l}_4$.  

Systematic errors can arise from several sources: finite volume
effects, neglecting of higher orders in $\chi$PT and finite lattice
spacing effects. These different corrections are accounted for
explicitly in eq.~(\ref{eq:mpfp}). Finite volume corrections are
described by the rescaling factors denoted by $F_{\mps^2}$ and
$F_{\fps}$, computed in the continuum theory. Notice that the
discretisation effects present in the neutral pion mass, see
section~\ref{subsec:isospin}, generate peculiar finite volume
corrections which have been recently analysed in
ref.~\cite{Colangelo:2010cu}. We shall comment on them later. We
investigated the effectiveness of one loop continuum $\chi$PT finite
volume corrections, as first computed in~\cite{Gasser:1986vb}, which
do not introduce any additional low energy constants. However, the
resummed expressions derived by Colangelo, D\"urr and Haefeli (CDH)
in~\cite{Colangelo:2005gd} describe the finite volume effects in our
simulations better, be it at the expense of the introduction of two
new free parameters, and are thus adopted for this analysis. To
$\mathcal{O}(\xi^2)$, these corrections read
\begin{eqnarray}
F_{\mps^2} &=& \left[1 - \sum_{n=1}^\infty\frac{\rho_n}{2\,
\lambda_n}\left(\xi\,I^{(2)}_m + \xi^2\,I^{(4)}_m\right) \right]^2 \nonumber\\
F_{\fps}   &=& 1 + \sum_{n=1}^\infty\frac{\rho_n}{\lambda_n}\left(\xi\,I^{(2)}_f 
+ \xi^2\,I^{(4)}_f\right)\, , \label{eq:cdh}
\end{eqnarray}
with geometric contributions defined as
\begin{eqnarray}
I^{(2)}_m &=& -2 K_1(\lambda_n)\nonumber \\
I^{(4)}_m &=& \left(\frac{101}{9} - \frac{13}{3}\,\pi + 8\,l_1 + \frac{16}{3}
\,l_2 - 5\,l_3 - 4\,l_4\right)K_1(\lambda_n) + \nonumber\\
          & & \left(-\frac{238}{9} + \frac{61}{6}\,\pi-\frac{16}{3}
\,l_1 - \frac{64}{3}\,l_2 \right)\frac{K_2(\lambda_n)}
{\lambda_n} \nonumber \\
I^{(2)}_f &=& -4 K_1(\lambda_n)\nonumber \\
I^{(4)}_f &=& \left(\frac{29}{18} - \frac{29}{12}\,\pi + 4\,l_1 + 
\frac{8}{3}\,l_2 - 6\,l_4\right)K_1(\lambda_n) + \nonumber \\
          & & \left(-\frac{307}{9} + \frac{391}{24}\,\pi-\frac{16}{3}\,l_1
 - \frac{64}{3}\,l_2 \right)\frac{K_2(\lambda_n)}
{\lambda_n}\, . \label{eq:geom_cont}
\end{eqnarray}
The $K_i$ are the modified Bessel functions and the low energy
constants $l_1$ and $l_2$ again receive renormalisation
corrections. Equations~(\ref{eq:cdh})~and~(\ref{eq:geom_cont}) use the
shorthand notation $\lambda_n = \sqrt{n} \mps L$.  
The $\rho_n$ in eq.~(\ref{eq:cdh}) are a set of multiplicities,
counting the number of ways $n^2$ can be distributed over three
spatial directions\footnote{These values are straightforwardly
  precomputed to any order, but are also given in,
  \textit{e.g.}~\cite{Colangelo:2005gd}.}. Because the finite volume
corrections in the case of the volumes used in the chiral fits are
fairly small to begin with and subsequent terms quickly decrease, the
sums over $n$ can be truncated rather aggressively without real loss
of precision. It is therefore unnecessary, in practise, to go beyond
the lowest contributions. The parameters $l_1$ and $l_2$, which are in
fact low energy constants appearing at NLO in $\chi$PT, cannot be
determined well from the small finite volume corrections alone. Priors
are therefore introduced as additional contributions to the $\chi^2$,
weighting the deviation of the parameters from their phenomenological
values by the uncertainties in the latter. 
The values used as priors are -0.4(6) for $\bar{l}_1$ and 4.3(1) for 
$\bar{l}_2$~\cite{Colangelo:2005gd}, as reported in table \ref{table:fitresults_c}. We used the 
largest available volumes for each ensemble, in the $\chi$PT fits. 
For those points, the difference between the finite volume and the infinite volume 
values estimated via CDH formulae for $\fps$ and $\mps^2$ are within $1\%$,
except for the runs B85.24 and A60.24 (see table~\ref{tab:par_sim} and table~\ref{tab:rawdata}), 
where they are about $1.5\%$ for both quantities.

Because of the automatic $\Oa$ improvement of the twisted mass action
at maximal twist, the leading order discretisation artefacts in the
chiral formulae of (\ref{eq:mpfp}) are at least of $\Oasq$, and  $\mathcal{O}(a^2\mu)$ for 
$\mps^2$. 
The mass and decay constant of the charged
  pion have been studied up to
  NLO~\cite{Munster:2003ba,Scorzato:2004da,Sharpe:2004ny} in the
  context of twisted mass chiral perturbation theory (tm$\chi$PT). The
  regime of quark masses and lattice spacings at which we have
  performed the simulations is such that $\mu_l \gtrsim a
  \Lambda_\mathrm{QCD}^2$. In the associated power counting, at
  maximal twist, the NLO tm$\chi$PT expressions for the charged pion
  mass and decay constant preserve their continuum form. The inclusion
  of the terms proportional to $D_{\mps^2,\fps}$, parametrising the lattice
  artifacts in eq.~(\ref{eq:mpfp}), represents an effective way of
  including sub-leading discretisation effects appearing at
  NNLO.
The finite lattice
spacing artefacts can of course not be determined using only data from
a single lattice spacing. In addition, including these terms when
analysing data with an insufficient range in $a$, may lead to mixing
of these degrees of freedom with continuum parameters and thereby
destabilise the fits. Hence, these terms were neglected for the
separate fits, but included to arrive at a qualitative estimate of
these systematic effects in a combined fit to the data at both lattice
spacings.

Finite size effects on our data at finite lattice spacing can be
analysed in the context of twisted mass chiral perturbation theory as
recently proposed in ref.~\cite{Colangelo:2010cu}.\,\footnote{Notice
  that, in principle, after performing the continuum limit at fixed physical volume,
  finite size effects can be analysed by means of continuum $\chi$PT.}
However, our present limited set of data with only a small number of
different volumes all of them at a single value of the lattice
spacing, is not sufficient to apply such an analysis. We plan,
however, to perform dedicated runs on different volumes to confront
our data to the finite size effect formulae of
ref.~\cite{Colangelo:2010cu} and to estimate in particular the size of
the pion mass splitting in this alternative way.

Finally, results from continuum $\chi$PT at NNLO can be included to examine the
effect of the truncation at NLO. They are given by 
\begin{eqnarray}
T_{\mps^2} &=& \frac{17}{102}\left(49 + 28\,l_1 + 32\,l_2 - 9\,l_3\right) + 4\,k_m \nonumber \\
T_{\fps} &=& -\frac{1}{6}\left(23 + 14\,l_1 + 16\,l_2 + 6\,l_3 - 6\,l_4\right) + 4\,k_f.
\end{eqnarray}
Two new parameters $k_m$ and $k_f$ enter these corrections.  Again, a
limited range of input pion masses may lead to poorly constrained
values of these newly introduced parameters, some degree of mixing
among different orders and fit instabilities. To retain predictive
power and stability, additional priors are given for $k_m$ and $k_f$,
both priors set to $0(1)$, analogously to what is done for
$l_1$ and $l_2$ in the CDH finite volume corrections.

To set the scale at each lattice spacing, we determine
$a\mu_\mathrm{phys}$, the value of $a\mu_l$ at which the ratio
$\sqrt{\mps^2 (L=\infty )}/\fps (L=\infty )$ assumes its physical
value. We can then use the value of $\fps$, or equivalently $\mps$, to
calculate the lattice spacing $a$ in fm from the corresponding
physical value. We also perform a chiral fit combining the two
different lattice spacings. With only two different values of $\beta$,
that are in fact fairly close to each other, a proper continuum limit
analysis cannot be performed. Instead, we treat this combined fit as a
check on the presence of lattice artefacts and the overall consistency
of the data. Without a scaling variable, such as the Sommer scale
$r_0$, the data from different lattice spacings cannot be directly
combined.  Rather, the ratios of lattice spacings and light quark mass
renormalisation constants ($Z_\mu = 1/Z_P$), as well as the
renormalised $B_0$ parameter are left free in the fit.

In order to estimate the statistical errors affecting our fitted
parameters, we generate at each of the $\mu_l$ values 1000 bootstrap
samples for $\mps$ and $\fps$ extracted from the bare correlators,
organised by blocks. For each sample, and combining
all masses, we fit $\mps^2$ and $\fps$ simultaneously as a function of
$\mu_l$. The parameter set from each of these fits is then a separate
bootstrap sample for the purposes of determining the error on our fit
results. By resampling $\fps$ and $\mps$ on a per-configuration basis,
correlations between these quantities are taken into account.

\begin{figure}[t]
  \centering
    \subfigure[\label{fig:amps2_vs_amu_190}]%
  {\includegraphics[width=0.46\linewidth]{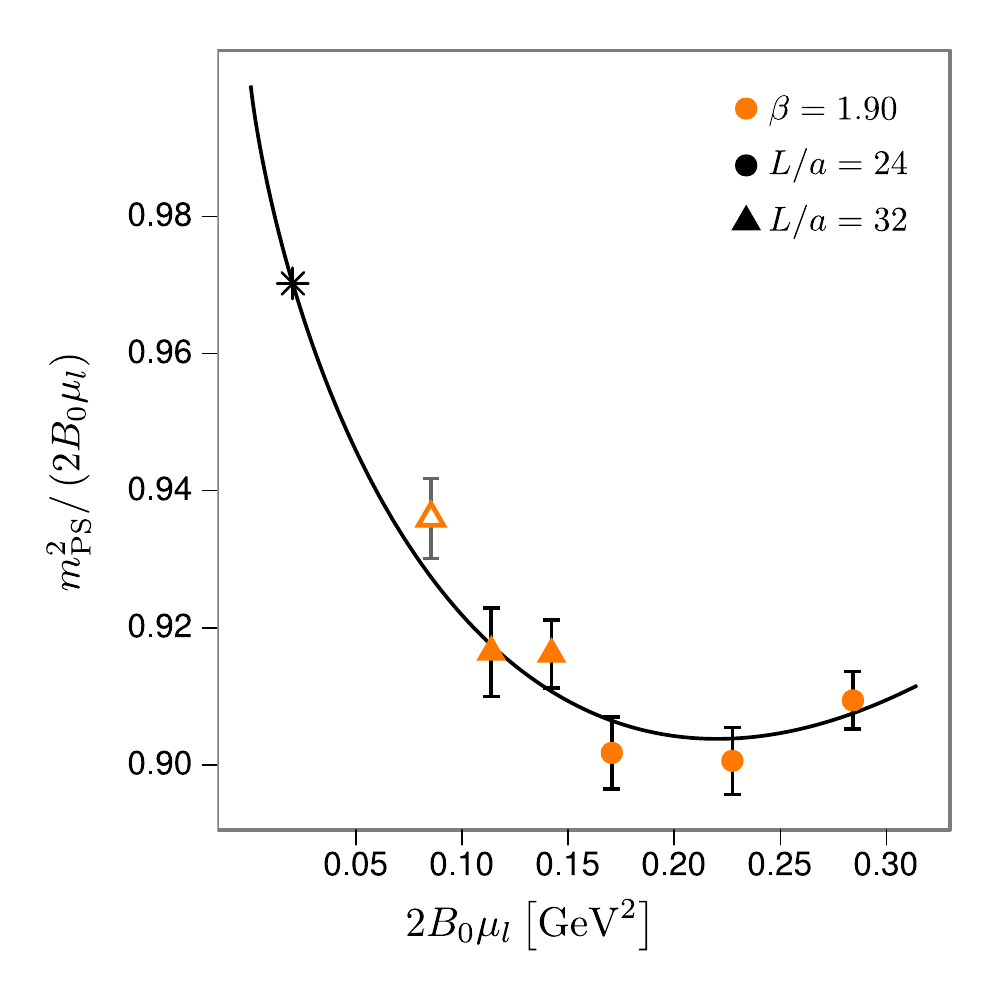}}
  \quad
  \subfigure[\label{fig:afps_vs_amu_190}]%
  {\includegraphics[width=0.44\linewidth]{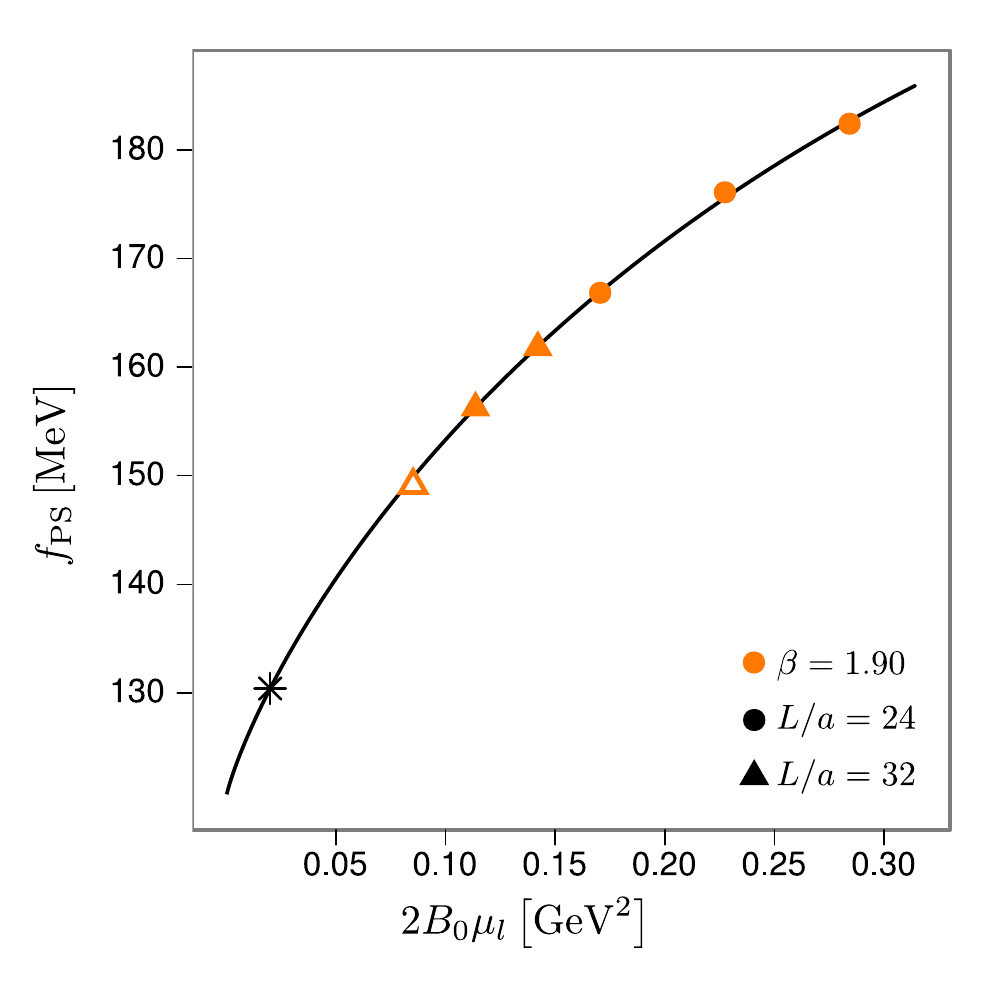}}
  \caption{
(a) The charged pseudoscalar mass ratio $\mps^2/2B_0\mu_l$ and (b) the pseudoscalar decay constant 
 $\fps$ as a function of $2B_0\mu_l$, 
for the ensemble at $\beta =1.90$, fitted to SU(2) chiral perturbation theory, eq.~(\ref{eq:mpfp}).
The scale is set by $a\mu_\mathrm{phys}$, the value of $a\mu_l$ at which the ratio
$\fps^{[L=\infty]}/\mps^{[L=\infty]}$ assumes its physical
value~\cite{Amsler:2008zzb} $f_\pi/m_\pi = 130.4(2)/135.0$ (black star).
The light twisted masses used in the fit range from $a\mu_l = 0.004$ to $0.010$. The lightest point 
(open symbol) lies outside our most conservative tuning criterion to maximal twist, and is not included in the fit.} 
\label{fig:chfit_190}
\end{figure}

\begin{figure}[t]
  \centering
    \subfigure[\label{fig:amps2_vs_amu_comb}]%
  {\includegraphics[width=0.46\linewidth]{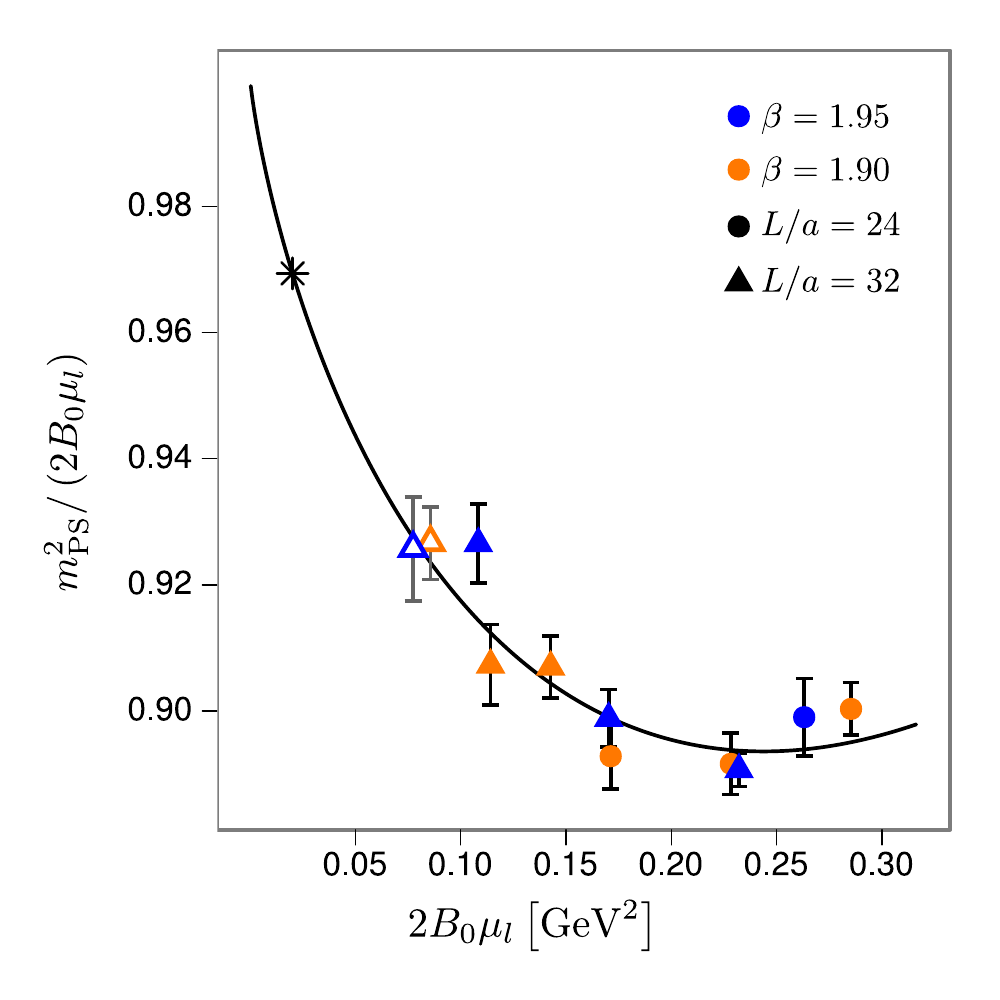}}
  \quad
  \subfigure[\label{fig:afps_vs_amu_comb}]%
  {\includegraphics[width=0.44\linewidth]{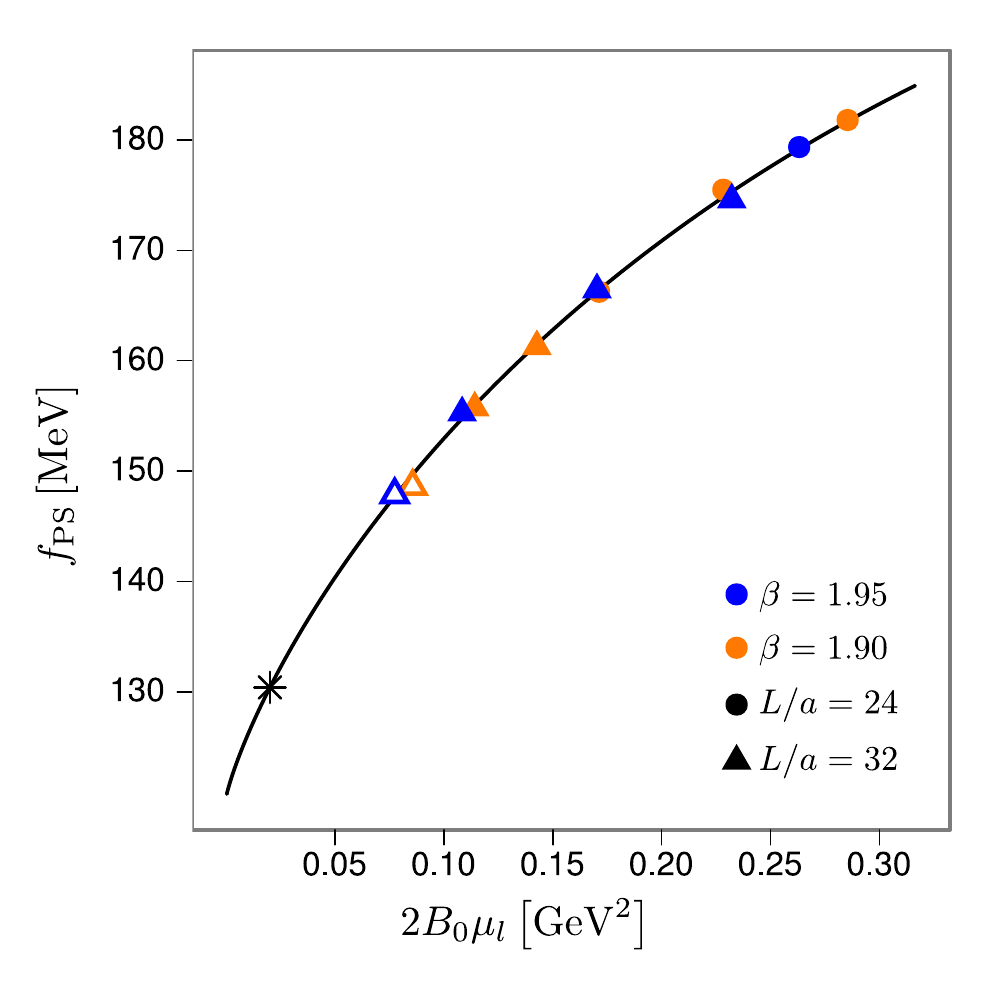}}
  \caption{
(a) The charged pseudoscalar mass ratio $(\mps /2B_0\mu_l)^2$ and (b) the 
pseudoscalar decay constant  $\fps$ as a function of $2B_0\mu_l$, for the 
combined ensembles at $\beta =1.90$ and $\beta =1.95$, and fitted to eq.~(\ref{eq:mpfp}). 
The scale is set as in figure~\ref{fig:chfit_190}  (black star).
The light twisted masses used in the fit range from $a\mu_l = 0.0035$ to $0.010$. 
The lightest point at $\beta =1.90$ (open orange symbol) and at $\beta =1.95$ 
(open blue symbol) lie outside our most conservative tuning criterion to maximal
 twist, and are not included in the fit.}
  \label{fig:chfit_comb}
\end{figure}

Our final results for the separate and combined fits are summarised in 
table~\ref{table:fitresults_c}. 
\begin{table}[t!]
  \centering
  \begin{tabular*}{.85\linewidth}{@{\extracolsep{\fill}}lrrrr}
    \hline\hline
    $\Bigl.\Bigr.$ & $\beta=1.90$ & $\beta=1.95$ & combined & priors\\
    \hline\hline
    $\bar{l}_3$      &  3.435(61)   &  3.698(73)  &  3.537(47)  & -\\
    $\bar{l}_4$      &  4.773(21)   &  4.673(25)  &  4.735(17)  & -\\
    $\bar{l}_1$      & -0.296(104)  & -0.430(93)  & -0.309(139) & -0.4(6)\\
    $\bar{l}_2$      &  4.260(12)   &  4.329(15)  &  4.325(10)  &  4.3(1)\\
    $f_0\ [\mathrm{MeV}]$   & 120.956(70)  & 121.144(83)  & 121.031(54)& -\\
    $f_\pi/f_0$             & 1.0781(18)   & 1.0764(18)   & 1.0774(17)& -\\
    $2B_0\mu_{u,d}/m_\pi^2$ & 1.029(16)    & 1.032(21)    & 1.030(13) & -\\
    $\langle r^2\rangle_s^\mathrm{NLO}\ [\mathrm{fm}^2]$
    & 0.7462(43) & 0.7237(51)  & 0.7375(34)  & - \\
    \hline
    $r^\chi_0/a(\beta=1.90)$   & 5.231(38)   & -           & 5.231(37)& -\\
    $r^\chi_0/a(\beta=1.95)$   & -           & 5.710(41)   & 5.710(42)& -\\
    $r^\chi_0(\beta=1.90)\ [\mathrm{fm}]$
    & 0.4491(43) & -          & 0.4505(40) & -\\
    $r^\chi_0(\beta=1.95)\ [\mathrm{fm}]$
    & -          & 0.4465(48) & 0.4439(39) & -\\
    $a(\beta=1.90)\ [\mathrm{fm}]$
    & 0.08585(53)& -          & 0.08612(42)& -\\
    $a(\beta=1.95)\ [\mathrm{fm}]$
    & -          & 0.07820(59)& 0.07775(39)& -\\
    \hline
  \end{tabular*}
  \caption{Results of the fits to SU(2) $\chi$PT for the ensembles at
    $\beta=1.95$ and $\beta=1.90$, separate and combined. The largest
    available volumes are used for each ensemble. Predicted quantities
    are: the low energy constants $\bar{l}_{3,4}$ (while
    $\bar{l}_{1,2}$ are introduced with priors), the charged
    pseudoscalar decay constant in the chiral limit $f_0$, the mass
    ratio $2B_0\mu_l/\mps^2$ at the physical point and the pion scalar
    radius $\langle r^2\rangle_s^\mathrm{NLO}$. 
 The scale is set by fixing the ratio
    $\fps^{[L=\infty]}/\mps^{[L=\infty]} = f_\pi/m_\pi = 130.4(2)/135.0$ to its physical
    value~\cite{Amsler:2008zzb}.
The chirally extrapolated Sommer
    parameter $r_0^\chi$ is determined separately and not included in
    the chiral fits.  For a comparison with the $N_{\rm f}=2$ ETMC
    results, see~\cite{Baron:2009wt}.}
  \label{table:fitresults_c}
\end{table}
The $\chi$PT fit ans\"atze provide a satisfactory description of the
lattice data, with a $\chi^2/{\rm d.o.f} = 5.68/3\simeq 1.9$ at
$\beta=1.95$, $\chi^2/{\rm d.o.f} = 4.31/5\simeq 0.9$ at $\beta=1.90$,
and $16.9/11\simeq 1.5$ for the combined fit.  We also predict the
scalar radius of the pion at next to leading order
\begin{equation}
\langle r^2\rangle_s^{\rm NLO} = \frac{12}{(4\pi f_0)^2}\left (\bar{l}_4 - \frac{13}{12}\right).
\end{equation}
The numerical values in table~\ref{table:fitresults_c} for the combined
fit show a very good agreement with the results from the separate
fits, and with errors at the percent level throughout. The fits for
$\fps$ and $\mps$ at $\beta =1.95$ are displayed in
figures~\ref{fig:chiralfit}(a) and (b), while in figures
~\ref{fig:chfit_190}(a) and (b) we show the analogous fits at $\beta
=1.90$.
Figures~\ref{fig:chfit_comb}(a) and (b) show the results for the fit
combining the two $\beta$ values.

The data presented here do not allow yet for a complete account of the
systematic effects, but we extract estimates of their magnitude by
extending the fits with additional terms as written down in
eq.~(\ref{eq:mpfp}). Checks were done for $\chi$PT NNLO terms and
$\Oasq$ corrections separately. Including NNLO corrections does not
lower the total $\chi^2$ of the fit, while we do observe a shift of several
 standard deviations for the lower order parameters already
present in the NLO fit. Using these shifted values to obtain the
implied NLO approximation produces fits with much larger values of
$\chi^2$. We conclude that the current data lack the precision and
range in quark masses to constrain NNLO effects,  the added degrees
of freedom mix with NLO effects and destabilise the fit instead. In
practise, we conclude that the systematic error from the
truncation of $\chi$PT is unobservable at the current level of
precision. Inclusion of $\Oasq$ corrections leads to similar observations,
as the difference between the lattice spacings and the statistical
accuracy of the data is too small to result in a stable fit. The fit
mixes $D_{\fps}$ and $D_{\mps^2}$ on the one hand and $f_0$, $B_0$ and
the rescaling in the lattice spacing and the quark mass on the other.

The chirally extrapolated Sommer scale $r_0^\chi$ has been determined
separately, using a fit of $r_0/a$ with quadratic dependence on the
bare light quark mass, as shown in figures \ref{fig:r0overa} and
\ref{fig:r0overa_2}, and using the lattice spacing determined by the
chiral fits. As also reported in table~\ref{table:fitresults_c}, the
obtained values are $r_0^\chi = 0.4491(43)$\,fm at $\beta =1.90$ and
$r_0^\chi = 0.4465(48)$\,fm at $\beta = 1.95$, where only statistical
errors are quoted.  For consistency, we also verified that a combined
chiral fit with the inclusion of $r_0/a$, as data points and
additional fit parameter, gives results anyway in agreement with the
strategy adopted here.

For our final estimates of the low energy constants $\bar{l}_{3,4}$
and the chiral value of the pseudoscalar decay constant $f_0$ we use
the predictions from the $\beta =1.95$ ensemble based on two important observations. First,
the strange quark mass in this ensemble is better tuned to the
physical value. Secondly a reduced
isospin breaking is observed at this finer lattice spacing.
The results for the $\beta =1.90$ ensemble and the
combined fits serve instead as an estimation of systematic uncertainties.
As a result of the current $N_f =2+1+1$ simulations we thus quote
\begin{equation}
\bar{l}_3 = 3.70(7)(26) ~~~~\bar{l}_4 = 4.67(3)(10)\, , 
\end{equation}
and $f_0 = 121.14(8)(19)$ MeV, where the first error comes from the
chiral fit at $\beta =1.95$, while the second quoted error
conservatively accommodates the central values from the $\beta =1.90$
and combined fits as a systematic uncertainty. The predictions for
$\bar{l}_3$ and $\bar{l}_4$ are in good agreement and with our
two-flavour predictions~\cite{Baron:2009wt} and with other recent
lattice determinations~\cite{Scholz:2009yz,Necco:2009cq}.

\section{Conclusions and Outlook}
\label{sec:concl}

In this paper we have presented the first results of lattice QCD
simulations with mass-degenerate up, down and mass-split strange and
charm dynamical quarks using Wilson twisted mass fermions at maximal
twist. This study constitutes a first step in our effort to describe
low energy strong dynamics and electroweak matrix elements by fully
taking into account the effects of a strange and a charm quark.

We have considered ensembles at slightly different lattice spacings
simulated with Iwasaki gauge action at $\beta =1.95$ with $a\approx
0.078$\,fm and $\beta = 1.90$ with $a\approx 0.086$\,fm.  The charged
pseudoscalar masses range from $270$ to $510$ MeV and we performed
fits to SU(2) chiral perturbation theory with all data at a value of
$\mps L\gtrsim 4$.  This analysis provides a prediction for the low
energy constants $\bar{l}_3 = 3.70(7)(26)$ and
$\bar{l}_4=4.67(3)(10)$, for the charged pseudoscalar decay constant
in the chiral limit $f_0 = 121.14(8)(19)$ MeV and for the scalar
radius at next-to-leading order $\langle r^2\rangle_s^{\rm NLO}=
0.724(5)(23)$ fm$^2$.  A companion paper~\cite{Baron:HL2010} describes
the less straightforward determination of the kaon and D-meson masses
for the same ensembles.

We have compared our results in the light meson sector with those
obtained for $N_{\rm f}=2$ flavours of maximally twisted mass
fermions, ref.~\cite{Baron:2009wt}. There, an extrapolation to the
continuum limit, a study of finite size effects and checks against
higher order $\chi$PT have been performed, leading to a controlled
determination of systematic errors.  The comparison we have carried
through does not show any significant difference between $N_{\rm f}=2$
and $N_{\rm f}=2+1+1$ flavours, at least at the present level of accuracy. 
These results would suggest that effects of the strange
and charm quarks are suppressed for these light observables, as it should be expected.
The same comparison has also been used for a first
investigation of lattice discretisation errors. As
figures~\ref{fig:discr} and~\ref{fig:discr2} show, the $N_{\rm f}=2+1+1$
data are completely consistent with the corresponding ones obtained
for $N_{\rm f}=2$, where the discretisation effects have turned out to
be very small. Thus, it can be expected that also for the case of
$N_{\rm f}=2+1+1$ flavours the lattice spacing effects will be small,
at least for the light meson sector considered here. 
Notice however that, at the present level of accuracy, there is still 
the possibility that cancellations occur  between physical contributions
 due to dynamical strange and charm quarks and lattice discretisation effects. 
A more accurate study at a significantly lower lattice spacing will 
allow to draw conclusions.

One aspect of the twisted mass formulation is the breaking of isospin
symmetry. Its effect is likely to be most pronounced in the lightest
sector, where lattice discretisation effects at $\Oasq$,
affecting the neutral pseudoscalar mass only, generate a mass splitting 
between the charged and the neutral pseudoscalar mesons.
While this mass splitting for
$N_f=2+1+1$ flavours has been found here to be larger than in the
$N_f=2$ simulations at fixed physical situation, we do not find
further effects in other quantities computed so far. This observation
is supported by theoretical arguments
\cite{Frezzotti:2007qv,Dimopoulos:2009qv} and consistent with our
experience in the $N_f=2$ flavour case.

We consider the present results to be encouraging to proceed with the
$N_f=2+1+1$ flavour research programme of ETMC. In particular, we want
to perform the non-perturbative renormalisation with dedicated runs
for $N_f=4$ mass-degenerate flavours, an activity which we have
started already. Furthermore, we want to compute the quark mass
dependence of many physical quantities towards the physical point
where the pion assumes its experimentally measured value.  We are
currently performing simulations at a significantly different and
lower lattice spacing than the present ensembles.  Both strategies,
smaller quark masses and smaller lattice spacings, will allow us to
estimate systematic effects on a quantitative level and to obtain in
this way accurate physical results in our $N_f=2+1+1$ flavour
simulations with statistical and systematical errors fully under
control.

\section*{Acknowledgements}

We want to thank the whole ETMC for a very fruitful and enjoyable collaboration. 
In particular, we gratefully acknowledge valuable suggestions and discussions 
with Beno\^it Blossier, Roberto Frezzotti, Andreas Nube, 
Giancarlo Rossi and Enno Scholz.

The computer time for this project was made available to us by the
John von Neumann-Institute for Computing (NIC) on the JUMP, Juropa and Jugene
systems in J\"ulich and apeNEXT system in Zeuthen, BG/P and BG/L in Groningen,
by BSC on Mare-Nostrum in Barcelona (www.bsc.es), and by the
computer resources made available by CNRS on the BlueGene system
at GENCI-IDRIS Grant 2009-052271 and CCIN2P3 in Lyon.  We thank
these computer centres and their staff for all technical advice
and help.

V.G. and D.P. thank the MICINN (Spain) for partial support 
under grant FPA2008-03373.
This work has been supported in part by the DFG
Sonder\-for\-schungs\-be\-reich/ Trans\-regio SFB/TR9-03 and the
EU Integrated Infrastructure Initiative Hadron Physics (I3HP)
under contract RII3-CT-2004-506078.  We also thank the DEISA
Consortium (co-funded by the EU, FP6 project 508830), for support
within the DEISA Extreme Computing Initiative.

\bibliographystyle{h-physrev5}
\bibliography{bibliography}

\end{document}